\documentclass[useAMS,usenatbib]{mn2e}

\usepackage{graphicx}

\title[
Imaging and Spectroscopy of NGC\,6369]{
Optical and Infrared Imaging and Spectroscopy of the Multiple-Shell 
Planetary Nebula NGC\,6369}

\author[G.\ Ramos-Larios, M.A.\ Guerrero, R.\ V\'azquez \& J.P.\ Phillips]{
G.\ Ramos-Larios$^{1}$\thanks{E-mail:
gerardo@astro.iam.udg.mx (GRL); 
mar@iaa.es (MAG); 
vazquez@astro.unam.mx (RV)}, 
M.A.\ Guerrero$^{2}$\footnotemark[1], 
R.\ V\'azquez$^{3}$\footnotemark[1] and 
J.P.\ Phillips$^{1}$\thanks{Deceased, April 29th, 2011}\\
$^{1}$Instituto de Astronom\'{\i}a y Meteorolog\'{\i}a, 
Av.\ Vallarta No.\ 2602, Col.\ Arcos Vallarta, C.P. 44130 Guadalajara, 
Jalisco, Mexico \\
$^{2}$Instituto de Astrof\'{\i}sica de Andaluc\'{\i}a, IAA-CSIC, 
C/ Glorieta de la Astronom\'{\i}a s/n, 18008 Granada, Spain \\
$^{3}$Instituto de Astronom\'{\i}a, Universidad Nacional Aut\'onoma de 
M\'exico, Apdo. Postal 877, 22800 Ensenada, B.C., Mexico}

\begin{document}

\date{Accepted 2011 October 25. Received 2011 October 25; in original form 2011 August 8}

\pagerange{\pageref{firstpage}--\pageref{lastpage}} \pubyear{2011}

\maketitle

\label{firstpage}

\begin{abstract}

NGC\,6369 is a double-shell planetary nebula (PN) consisting of a bright 
annular inner shell with faint bipolar extensions and a filamentary envelope.    
We have used ground- and space-based narrow-band optical and near-IR images, 
broad-band mid-IR images, optical long-slit echelle spectra, and mid-IR 
spectra to investigate its physical structure.  
These observations indicate that the inner shell of NGC\,6369 can 
be described as a barrel-like structure shape with polar bubble-like 
protrusions, and reveal evidence for H$_2$ and strong polycyclic aromatic hydrocarbons 
(PAHs) emission from a photo-dissociative region (PDR) with molecular 
inclusions located outside the bright inner shell.  
High-resolution \emph{HST} narrow-band images reveal an intricate 
excitation structure of the inner shell and a system of ``cometary'' knots.  
The knotty appearance of the envelope, the lack of kinematical evidence for 
shell expansion and the apparent presence of emission from ionized material 
outside the PDR makes us suggest that the envelope of NGC\,6369 is not a 
real shell, but a flattened structure at its equatorial regions.  
We report the discovery of irregular knots and blobs 
of diffuse emission in low-excitation and molecular line emission that 
are located up to 80\arcsec\ from the central star, well outside the 
main nebular shells.  
We also show that the filaments associated to the polar protrusions have spatial extents consistent with post-shock 
cooling regimes, and likely represent regions of interaction of these structures 
with surrounding material.
\end{abstract}


\section{Introduction}

The planetary nebula (PN) NGC\,6369 possesses a double-shell morphology 
consisting of a round bright inner shell and a fainter filamentary outer 
shell or envelope \citep{SCM92}.  
Its most remarkable morphological features are two disparate 
extensions of the inner shell along the east and west directions; 
the latter appears as a large, filamentary blister or \emph{ansae}, 
whilst the former is a bifurcated structure.  

Using low-dispersion spectroscopy, \citet{Monteiro_etal04} modelled 
the peculiar morphology of NGC\,6369 as a tilted collar of gas and 
two symmetrically located caps of emission.
This interpretation is consistent with the spheroidal shell and faint 
bipolar extensions proposed by \citet{SL06} based on a single long-slit 
high-dispersion spectrum obtained along the nebular major axis.  
These simplified models do not deal with the bifurcated structure 
to the eastern side of the inner shell, nor they consider 
the filamentary appearance of the envelope.  
These have been proposed to represent fast low ionisation emission regions 
\citep[FLIERs,][]{Hajian_etal97}, although evidence for fast expansion 
velocities is lacking.

\begin{figure*}
\centering
\includegraphics[width=2.0\columnwidth]{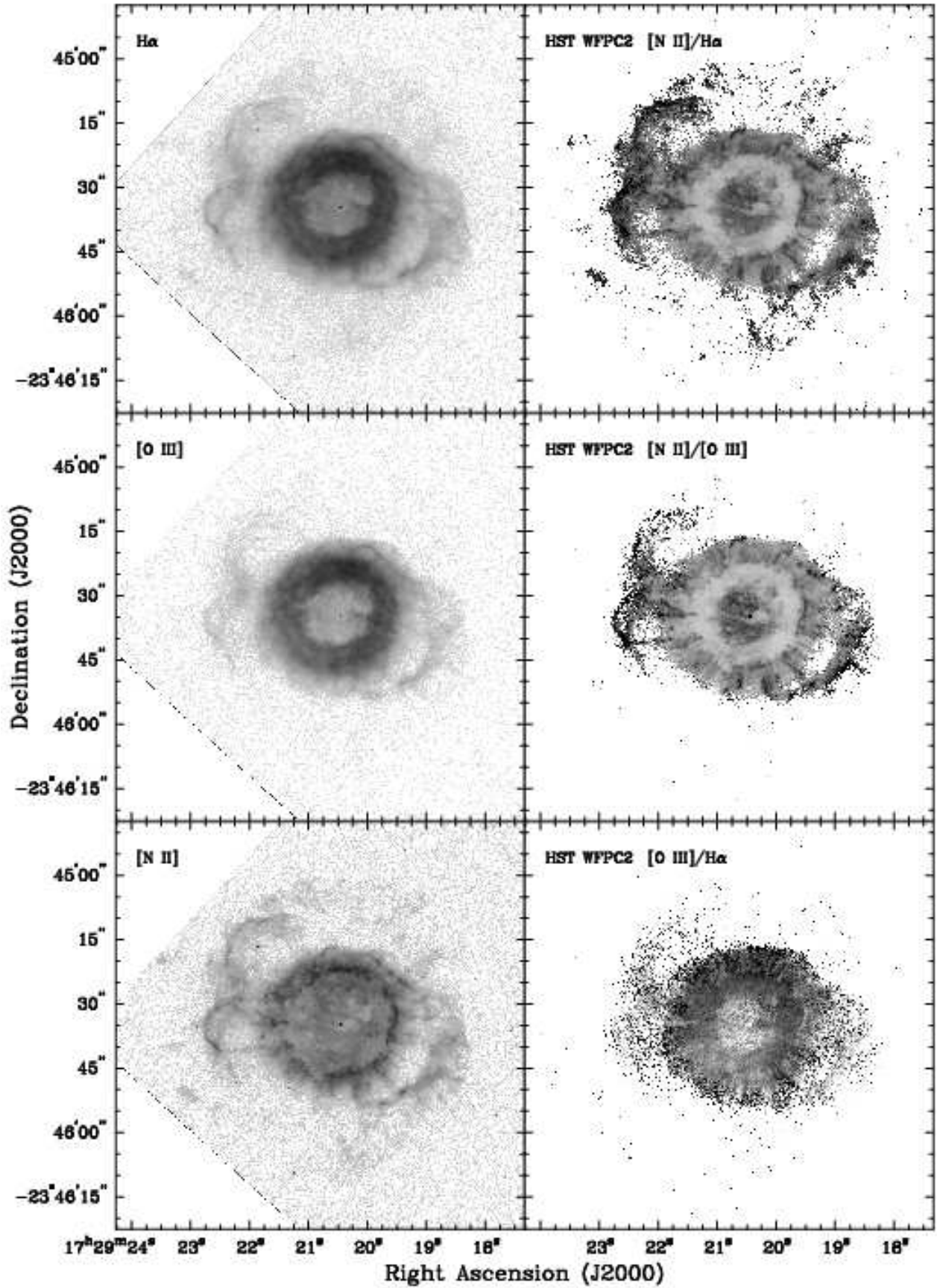}
\caption{
({\it left panels}) 
\emph{HST} WFPC2-WF3 images of the inner shell of NGC\,6369 in the 
H$\alpha$, [O~{\sc iii}] $\lambda$5007, and [N~{\sc ii}] $\lambda$6583 
emission lines, and 
({\it right panels}) 
[N~{\sc ii}]/H$\alpha$, [N~{\sc ii}]/[O~{\sc iii}], and 
[O~{\sc iii}]/H$\alpha$ ratio maps.  
}
\label{HST.img}
\end{figure*}

The H~{\sc i} and He~{\sc ii} Zanstra temperatures of NGC\,6369 have been 
reported to be $T_{\rm Z}$(H~{\sc i}) $\cong T_{\rm Z}$(He~{\sc ii}) $\cong$ 
70,000~K \citep[e.g.,][]{Phillips03}, comparable to estimates deriving 
from the energy balance method \citep{PB-S08}.  
The similarity of the H~{\sc i} and He~{\sc ii} Zanstra temperatures 
suggests that the nebula is optically thick to H ionizing radiation, 
and thus that significant amounts of molecular material and dust may 
be present outside the ionisation bound optical nebula.  
This is consistent with the \emph{ISO} (Infrared Space Observatory) 
indications of emission from polycyclic aromatic hydrocarbons (PAHs) 
bands \citep{CB05} that would be produced at a photo-dissociation 
region (PDR) located outside of the ionized shell.  
There is also evidence of dust emission from NGC\,6369.  
\emph{ISO} spectroscopic observations of NGC\,6369 have showed it to 
be one of the few evolved PNe to possess the 21 and 30 $\mu$m emission 
features \citep{HWT01} that are typically observed in C-rich AGB stars 
and PNe \citep[e.g.,][]{KVH99,VWR00,HVK09}. The origin of these spectral features is unclear, although it has been 
suggested that the 21 $\mu$m emission may arise from an extraordinarily 
wide range of agents, including hydrogenated fullerenes, nanodiamonds, 
SiC+SiO$_2$ grains, and TiC nanoclusters \citep[see e.g.,][ and references therein]{ZJL09}. The carbon dust origin in NGC\,6369 is consistent with its high C/O ratio, $\sim$2 \citep{ZA86,AK87}.

In order to investigate the kinematics and three-dimensional structure of 
the main nebula and envelope of NGC\,6369 and to confirm the occurrence of 
an outer envelope of molecular material, we have acquired new narrow-band 
near-IR H$_2$ and optical images, and long slit high-dispersion spectroscopic 
observations that have been examined in conjunction with archival mid-IR 
\emph{ISO} spectra and \emph{Spitzer} images and spectra, as well as optical 
narrow-band \emph{HST} images.  
The observations and archival data are presented in \S\ref{sec_obs}, and 
the results on the morphology, and mid-IR spectroscopy are 
described in \S\ref{sec_mor}, and \S\ref{mid_ir_sec}, 
respectively.  
The kinematics and the discussion are presented in \S\ref{sec_dis} and final conclusions are presented in \S\ref{sec_con}.

\section[]{Observations}\label{sec_obs}

\subsection{Optical imaging}

Narrow-band images in the [N~{\sc ii}], H$\alpha$, and [O~{\sc iii}] 
emission lines were obtained on September 1, 2008 using ALFOSC (Andalucia 
Faint Object Spectrograph and Camera) at the 2.56-m Nordic Optical 
Telescope (\emph{NOT}) in the Observatorio de Roque de los Muchachos 
(ORM, La Palma, Spain).  
The detector was a 2048$\times$2048 EEV CCD with plate scale 
0$\farcs$19 arcsec~pix$^{-1}$, resulting in a field of view 
(FoV) of 6$\farcm$5$\times$6$\farcm$5.  
The central wavelengths and bandwidths of the filters are 
$\lambda_{c}$=6589\,\AA\ and $\Delta\lambda$=9\,\AA\ for [N~{\sc ii}], 
$\lambda_{c}$=6568\,\AA\ and $\Delta\lambda$=8\,\AA\ for H$\alpha$, and 
$\lambda_{c}$=5007\,\AA\ and $\Delta\lambda$=8\,\AA\ for [O~{\sc iii}].    
The spatial resolution achieved during the observations, as 
determined from the FWHM of stars in the FoV, was 0$\farcs$95. 
Either two or three individual frames with integration times of 300\,s were taken for each filter, leading to total exposure times of 900\,s for the [N~{\sc ii}] image, and 600\,s for the H$\alpha$ and [O~{\sc iii}] 
images.  
The data were bias-subtracted and flat-fielded with twilight flats 
using standard IRAF\footnote{
IRAF, the Image Reduction and Analysis Facility, is distributed by 
the National Optical Astronomy Observatory, which is operated by 
the Association of Universities for Research in Astronomy (AURA) 
under cooperative agreement with the National Science Foundation.} 
routines.

\begin{figure*}
\centering
\includegraphics[width=1.0\columnwidth]{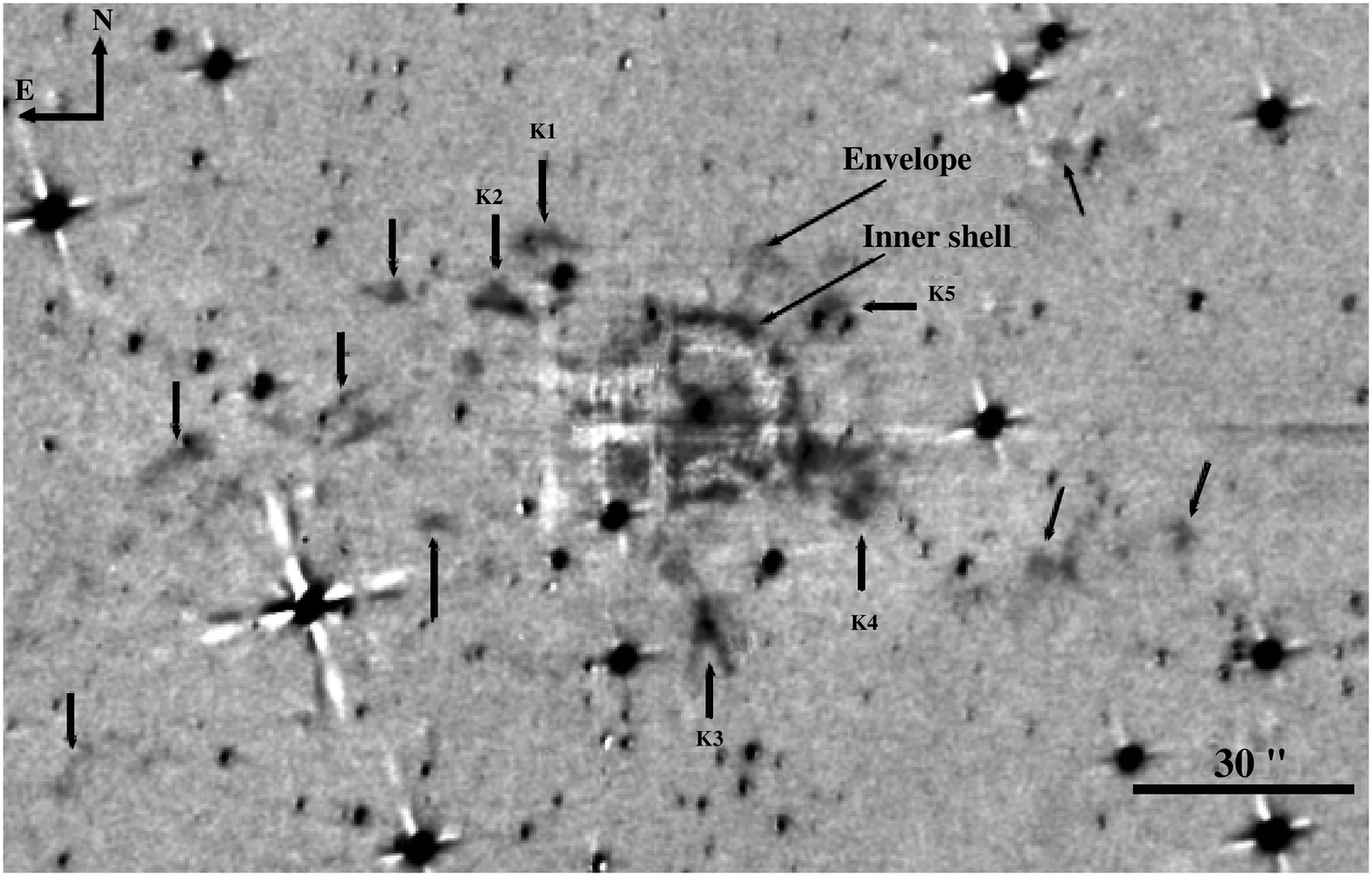}
\includegraphics[width=1.0\columnwidth]{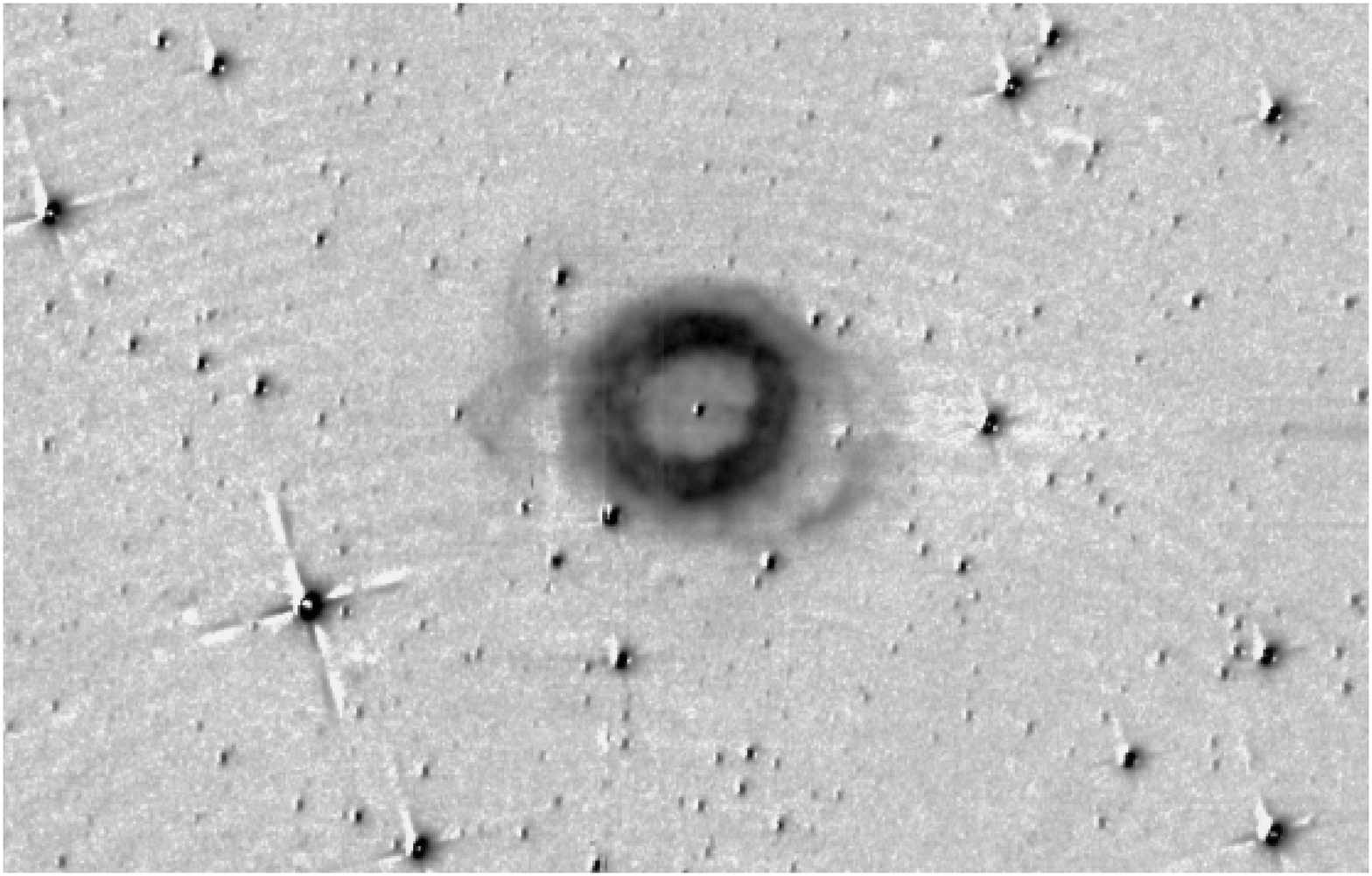}
\caption{
WHT LIRIS H$_2$ ({\it left}) and Br$\gamma$ ({\it right}) background-subtracted images of NGC\,6369.   
The dark and light diffraction spikes are a result of the rotation 
of the focal reducer during the acquisition of the images.  
The envelope and inner shell are labeled.  
The arrows point to knots and blobs of diffuse H$_2$ emission.  
In the envelope, five knots, marked as knots K1 to K5, have 
been highlighted.  
}
\label{H2.img}
\end{figure*}

Narrow-band \emph{HST} WFPC2 images of NGC\,6369 in the [N~{\sc ii}] 
$\lambda$6583, H$\alpha$, and [O~{\sc iii}] $\lambda$5007 emission 
lines were retrieved from MAST, the Multimission Archive at the Space 
Telescope Science Institute\footnote{
STScI is operated by the Association of Universities for Research in 
Astronomy, Inc., under NASA contract NAS5-26555.} (Prop.\ ID: 9582, Hubble Heritage Observations of NGC\,6369, PI: Keith Noll).  
The nebula was registered by the WF3 detector with a 
pixel scale of 0\farcs09.  
Four individual exposures were obtained using the 
F658N (pivot wavelength $\lambda_{p}$=6590.8\,\AA, $\Delta\lambda$=28.5\,\AA), 
F656N ($\lambda_{p}$=6563.8\,\AA, $\Delta\lambda$=21.5\,\AA), and 
F502N ($\lambda_{p}$=5012.4\,\AA, $\Delta\lambda$=26.9\,\AA) filters, 
resulting in total exposure times of 1,040\,s for the [N~{\sc ii}] 
$\lambda$6583 image, and 640\,s for the H$\alpha$ and [O~{\sc iii}] 
$\lambda$5007 images (Figure~\ref{HST.img}).

\subsection{Near-IR imaging}

Narrow-band H$_2$, Br$\gamma$ and continuum $K_c$ near-IR images of 
NGC\,6369 were obtained on 2010 June 27 using LIRIS, the Long-Slit 
Intermediate Resolution Infrared Spectrograph \citep{Pulido_etal03}, 
at the Cassegrain focus of the 4.2-m William Herschel Telescope 
(\emph{WHT}) on Roque de Los Muchachos Observatory (ORM, La Palma, Spain).  
The detector was a 1k$\times$1k HAWAII array sensitive 
in the spectral range from 0.8 to 2.5 $\mu$m.  
The plate scale is 0$\farcs$25 pix$^{-1}$ and the FoV 
4$\farcm$27$\times$4$\farcm$27.  
The narrow-band filters have $\lambda_c$=2.1218 $\mu$m and 
$\Delta\lambda$=0.032 $\mu$m for H${_2}$~(1$-$0)~S(1), 
$\lambda_c$=2.1658 $\mu$m and $\Delta\lambda$=0.032 $\mu$m for Br$\gamma$, and 
$\lambda_c$=2.270 $\mu$m and $\Delta\lambda$=0.034 $\mu$m for $K_c$.  

Ten 60\,s exposures were taken for each filter, for total effective 
exposure times of 600\,s.  
The jitter option was used to increase the sampling of the sky: 
the telescope pointing of each exposure is shifted by a few 
pixels, thus rastering the nebula to different locations on the 
detector. 
Each series of observations on the object was followed by similar series 
of observations on adjacent blank sky positions.  
The reduction of the LIRIS data was carried out using the dedicated 
software LIRISDR (LIRIS Data Reduction package), a pipeline for the 
automatic reduction of near-IR data developed within the IRAF 
environment.  
The reduction performed by LIRISDR includes standard and additional 
non-standard steps such as bad pixel mapping, cross-talk correction, 
flat-fielding, sky subtraction, removal of reset anomaly effect, 
field distortion correction, and final image co-addition. 
The continuum contribution to the H$_2$ and Br$\gamma$ emission was 
subsequently removed using the $K_c$ image.  
In this process, we did not find necessary to scale the images since 
the H$_2$, Br$\gamma$, and $K_c$ filters have very similar bandwidths 
and transmissions.  
The spatial resolution on the WHT images, as determined from the FWHM 
of stars in the field of view, is $\sim$0$\farcs$8. 
The narrow-band continuum-subtracted H$_2$ and Br$\gamma$ images 
of NGC\,6369 are presented in Figure~\ref{H2.img}.

\subsection{Mid-IR imaging}

\emph{Spitzer} Infrared Array Camera \citep[IRAC;][]{Fazio_etal04} 
images of NGC\,6369 (Figure 3) were retrieved from the NASA/IPAC Infrared 
Science Archive (IRSA).  
The images, obtained on 2005 September 22 as part of Program 20119 (The 
Darkest Cloud, An IRAC/MIPS Survey of the Pipe Nebula, PI: C.J.\ Lada), 
have been processed as described in the IRAC data handbook\footnote{
http://ssc.spitzer.caltech.edu/irac/dh/iracdata\_handbook\_ 3.0.pdf.}.  
The observations employed filters having isophotal wavelengths (and 
bandwidths $\Delta\lambda$) of 
3.550 $\mu$m ($\Delta\lambda$=0.75 $\mu$m), 
4.493 $\mu$m ($\Delta\lambda$=1.9015 $\mu$m), 
5.731 $\mu$m ($\Delta\lambda$ = 1.425 $\mu$m), 
and 7.872 $\mu$m ($\Delta\lambda$ = 2.905 $\mu$m).  
IRAC spatial resolution varies between $\simeq$1\farcs7 and $\simeq$2\farcs0 
\citep{Fazio_etal04}, and is reasonably similar in all IRAC bands, although 
the diffraction halo at 8 $\mu$m is stronger than at the other bands.

\begin{figure*}
\begin{center}
\includegraphics[width=\columnwidth]{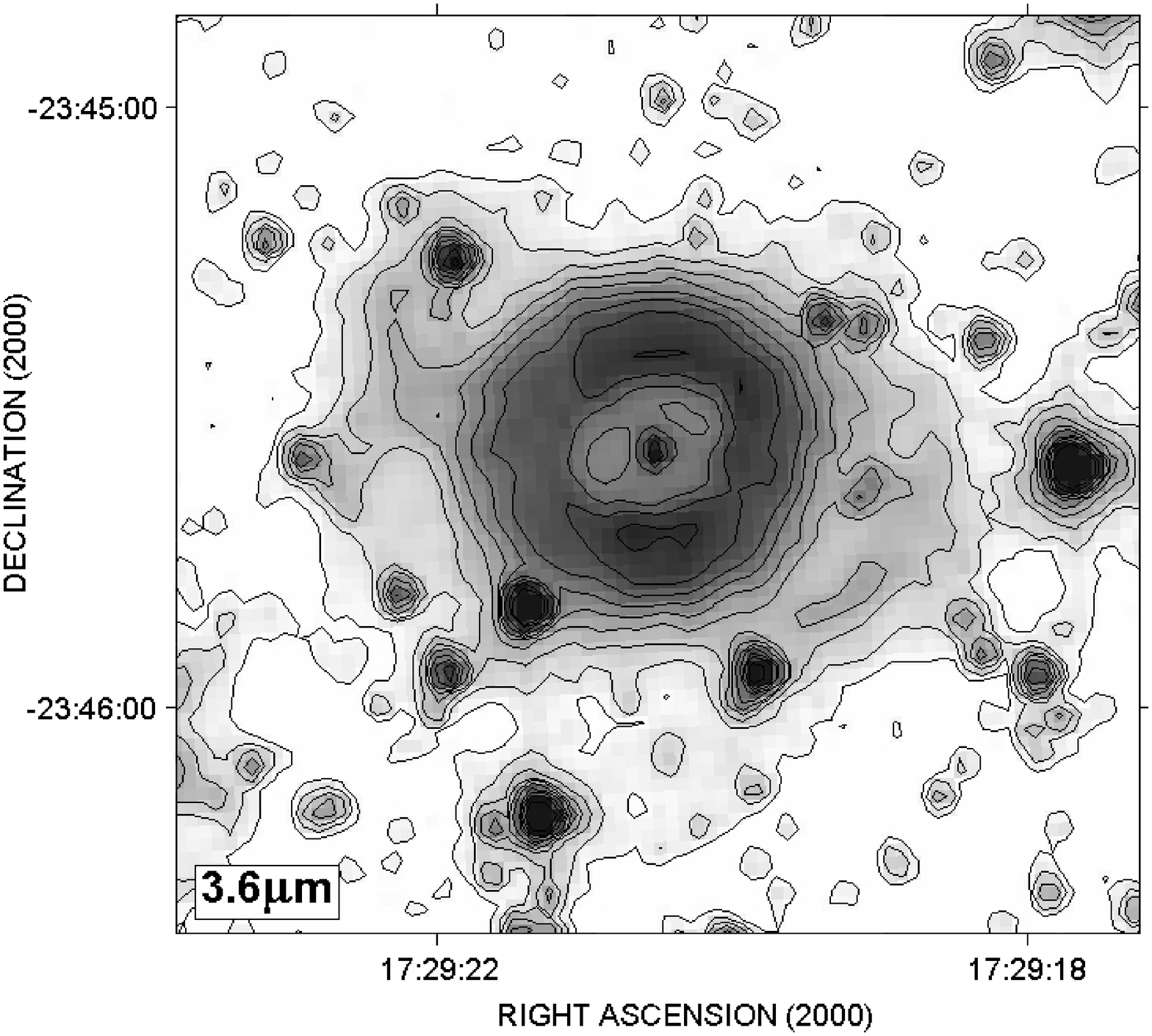}
\includegraphics[width=\columnwidth]{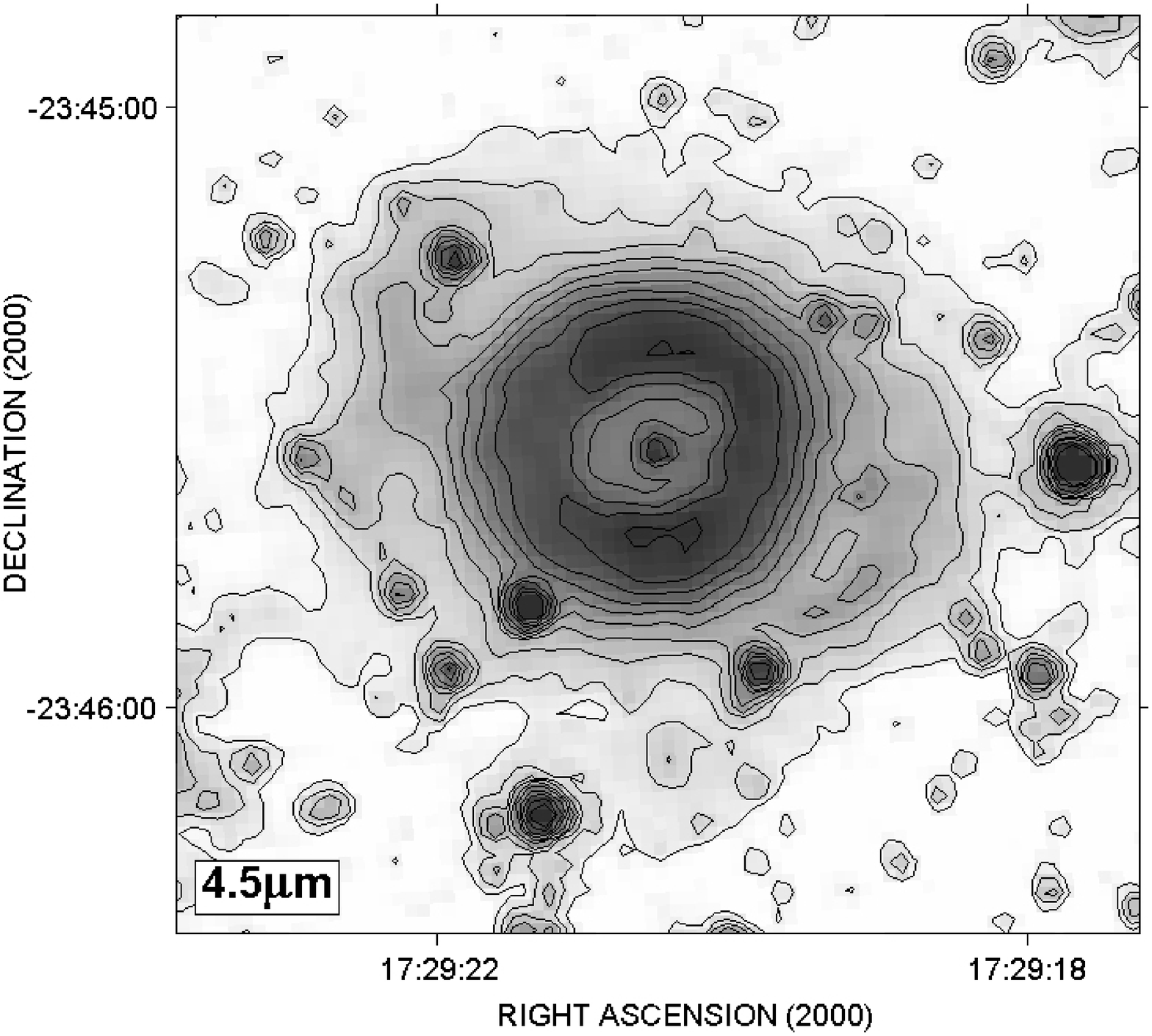}
\includegraphics[width=\columnwidth]{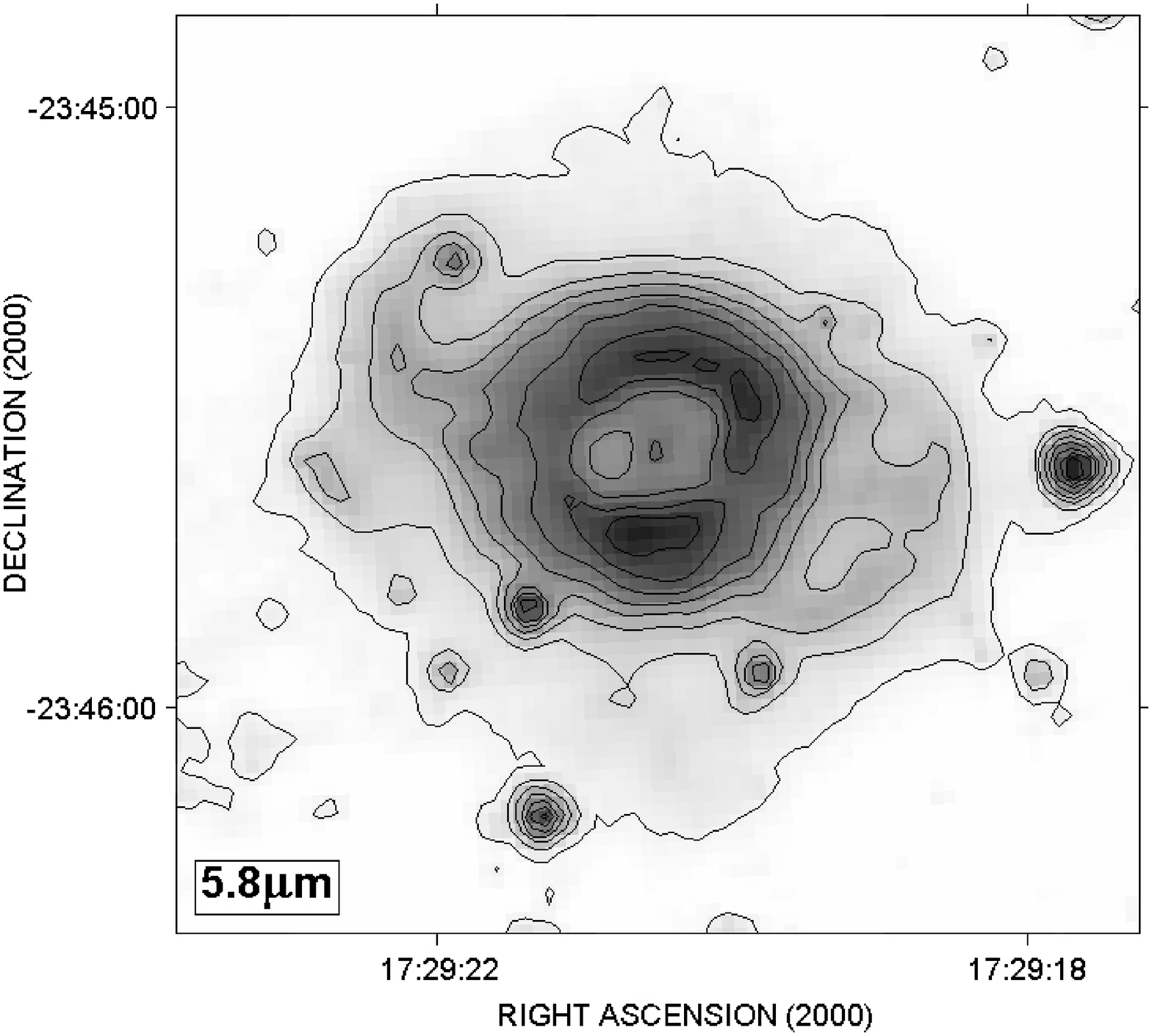}
\includegraphics[width=\columnwidth]{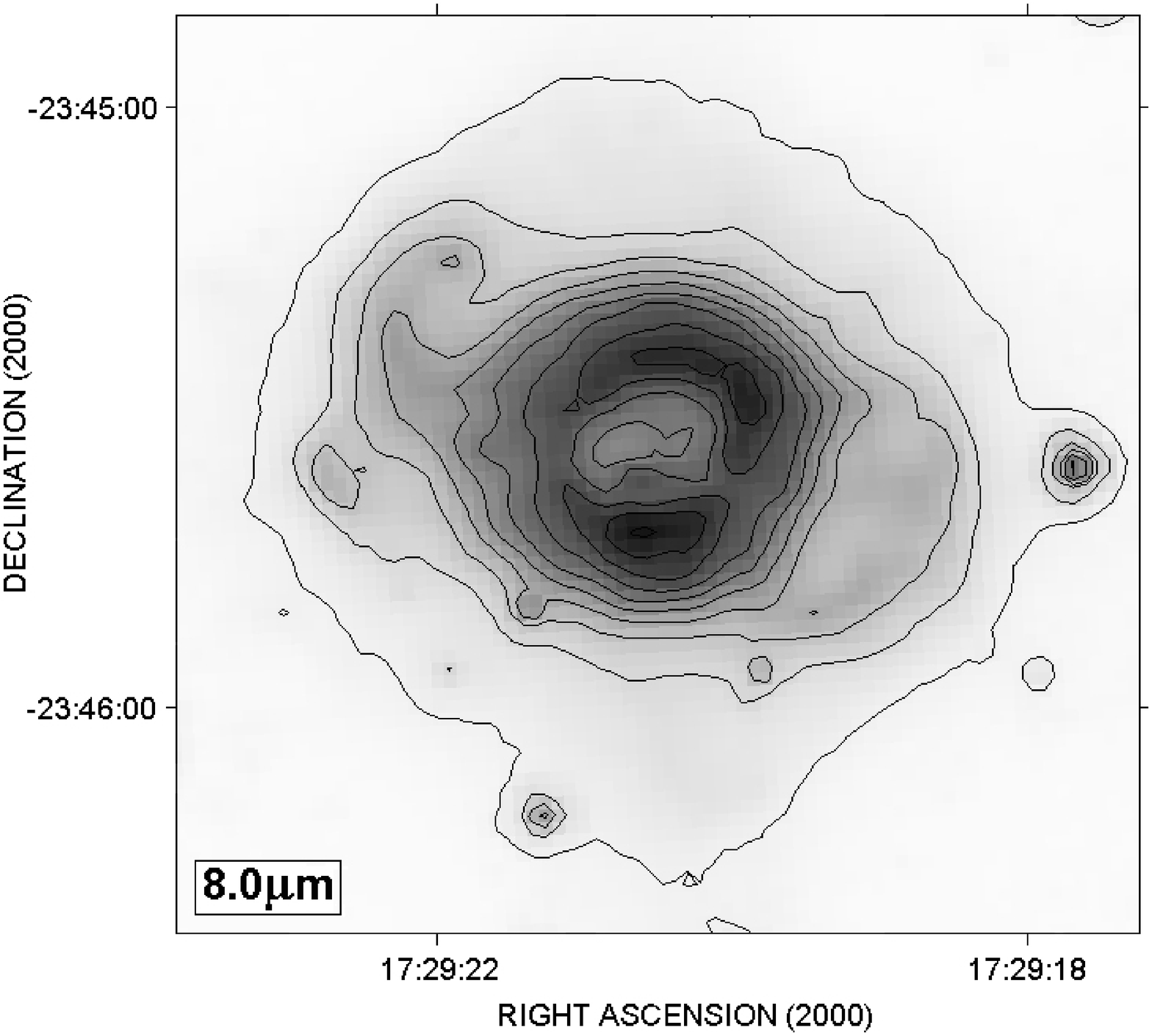}
\caption{
\emph{Spitzer} IRAC images of NGC\,6369 overplotted by contours.  
The contours correspond to levels of the intrinsic surface brightness, 
$SB$, defined by $SB_n$=A$\times$10$^{(n-1)\,C} -B$ MJy sr$^{-1}$, where 
$n$ is the contour level (n=1 corresponds to the lowest and outermost 
level), $B$ is the background level, and $A$ and $C$ are scale parameters. 
The contour parameters [$A$, $B$, $C$] are given by (1.6, 1.018, 0.199) 
at 3.6 $\mu$m, (1.45, 0.792, 0.196) at 4.5 $\mu$m, (7.3, 5.264, 0.17) at 5.8 $\mu$m, 
and (20.5, 15.341, 0.173) at 8.0 $\mu$m. 
We note that the blob of diffuse emission in the 8 $\mu$m image 
inside the inner shell is an artifact, an echo of the brightest 
region of the rim.  
}
\label{irac_maps}
\end{center}
\end{figure*}

\subsection{Long-slit high-dispersion optical spectroscopy}

Long-slit high dispersion optical spectroscopy of NGC\,6369 
was obtained on July 17-19 2008 using the Manchester Echelle 
Spectrometer \citep[MES,][]{Meaburn_etal03} mounted on the 
2.1\,m (f/7.5) telescope at the Observatorio Astron\'omico 
Nacional de San Pedro M\'artir (OAN-SPM, Mexico).  
A 1k$\times$1k SITe CCD was used as detector with a 2$\times$2 on-chip 
binning, resulting on a plate scale of 0\farcs6\,pix$^{-1}$ and a 
dispersion of 0.1\,{\AA}~pix$^{-1}$.  
Since MES has no cross dispersion, a $\Delta\lambda$=90\,\AA\ bandwidth 
filter was used to isolate the 87$^{\rm th}$ order covering the spectral 
range which includes the H$\alpha$ and [N~{\sc ii}] $\lambda6583$ lines. 
At this spectral order, the slit width of 150-$\mu$m (2\arcsec) that 
was set during the observations corresponds to a spectral resolution 
of $\simeq12$\,km\,s$^{-1}$. 

Five long-slit spectra, labeled from 1 to 5 on Figure~\ref{slits.img}, 
were obtained to map the kinematics at different regions of the nebula. 
The position angles (PAs) for slits 1 to 5 are $+74\degr$, 
$+80\degr$, $-22\degr$, $+22\degr$, and $-86\degr$, respectively.
The exposure time was 1,200\,s for the spectrum at PA=+80{\degr} (slit \# 2) and 1,800\,s for the rest. 
The seeing during the observations, as determined from the 
FWHM of stars in the FoV, varied from 1\farcs2 to 2\farcs0. 
The spectra were wavelength calibrated with a Th-Ar arc lamp to an accuracy 
of $\pm1$\,km\,s$^{-1}$.

\subsection{Mid-IR spectroscopy}

\emph{Spitzer} Infrared Spectrograph \citep[IRS;][]{Houck_etal04} 
mid-IR spectroscopic observations of NGC\,6369 obtained through the 
\emph{Spitzer} Program 40115 (Dual Dust Chemistry in Wolf-Rayet Planetary 
Nebulae, PI: G.\ Fazio) were retrieved from IRSA.  
The observations consisted of short (5.1-8.5 $\mu$m) and long (7.4-14.2 
$\mu$m) wavelength spectra taken with the \emph{Spitzer} IRS Short-Low 
module 2 (SL2) and 1 (SL1), respectively.  
Both modules have rectangular apertures $\sim$3\farcs6$\times$57\arcsec\ 
that, for these particular observations, were oriented along PA=177.18$\degr$.  
The observations were acquired at closely similar positions 
$\alpha$(J2000.0)=17$^{\rm h}$\,29$^{\rm m}$\,20\fs37 and 
$\delta$(J2000.0)=--23\degr\,45\arcmin\,24\farcs55 (positions A \& C), and 
$\alpha$(J2000.0)=17$^{\rm h}$\,29$^{\rm m}$\,20\fs44 and 
$\delta$(J2000.0)=--23\degr\,45\arcmin\,43\farcs55 (positions B \& D), 
resulting in overlapping slits covering the nebula along its minor axis.  
The spectral resolution of the observations, $\lambda$/$\Delta\lambda$, 
varied between 64 and 128.  
The \emph{Spitzer} SL1 \& SL2 spectra of the source are illustrated in 
Figure~\ref{IRS}, where we show the results for overlapping apertures 
across the centre of the source (labelled A to D).
All positions cover the inner shell, but positions A and C map 
the north portion of the envelope and positions B and D the 
south portion. 

Similarly, we made use of mid-IR \emph{ISO}\footnote{
Based on observations with ISO, an ESA project with instruments funded 
by ESA Member States (especially the PI countries: France, Germany, the 
Netherlands and the United Kingdom) and with the participation of ISAS 
and NASA} 
Short Wave Spectrometer \citep[SWS;][]{dG_etal96} spectroscopic 
observations of NGC\,6369 acquired on 1997 February 14 as part of 
program 45601901 (PI: S. G\'orny).  
\emph{ISO} SWS provides spectra in the 
short (2.38-12.1 $\mu$m), 
medium (12-29 $\mu$m), and 
long (29-45.3 $\mu$m) wavelength ranges, 
acquired through apertures with sizes 
14\arcsec$\times$20\arcsec, 
14\arcsec$\times$27\arcsec, and 
20\arcsec$\times$22\arcsec, respectively.  
The \emph{ISO} SWS spectra have an exposure time of 1,140\,s, and were centred on 
$\alpha$(J2000.0)=17$^{\rm h}$\,29$^{\rm m}$\,20\fs78 and 
$\delta$(J2000.0)=--23\degr\,45\arcmin\,32\farcs.1.

\section{Morphology}\label{sec_mor}

An inspection of the optical and infrared images of NGC\,6369 
(Figures~\ref{HST.img}, \ref{H2.img} and \ref{irac_maps}) 
confirms the three morphological components previously described 
in the main nebula: a bright inner shell, two extensions, and an 
envelope.  
Faint blobs of diffuse emission are also detected outside 
the main nebular shells.  
Next we describe in detail the morphology of the main nebular 
shell and this newly detected extended emission.

\subsection{The main nebula}\label{neb_morp_sec}

\subsubsection{Optical morphology}

The \emph{HST} WF3 H$\alpha$ and [O~{\sc iii}] images of the inner shell 
and envelope of NGC\,6369 presented in Figure~\ref{HST.img} show a thick 
($\sim$5\farcs5) annular shell with inner radius $\simeq$8\farcs3.  
The surface brightness is noticeably uneven, with patches of 
bright and diminished emission.  
Several dark knots and lanes, reminiscent of those seen in other 
PNe \citep[e.g., IC\,4406 and NGC\,6720,][]{O'Dell_etal2002} are 
also detectable.
 
The [N~{\sc ii}] image reveals a dramatically different view of the inner 
shell, with the H$\alpha$ and [O~{\sc iii}] dark knots and lanes turning 
into bright [N~{\sc ii}] spots and filaments, and an intricate system of 
[N~{\sc ii}] filaments filling the innermost regions of the shell.  
Radial filaments emanate outwards from most of the [N~{\sc ii}] knots, 
which are thus revealed as ``tadpole'' or ``cometary'' features very 
similar to those seen, e.g., in NGC\,2392 \citep{O'Dell_etal2002} and 
NGC\,7354 \citep{Contreras_etal2010}.  
Interestingly, there is a bright ring of these cometary knots, at the 
location of the outer rim of the bright H$\alpha$ and [O~{\sc iii}] 
shells at a distance of $\simeq$12\farcs4 of the central star, whose 
tails extend at least up to $\sim$20\arcsec.

\begin{figure}
\begin{center}
\includegraphics[width=\columnwidth]{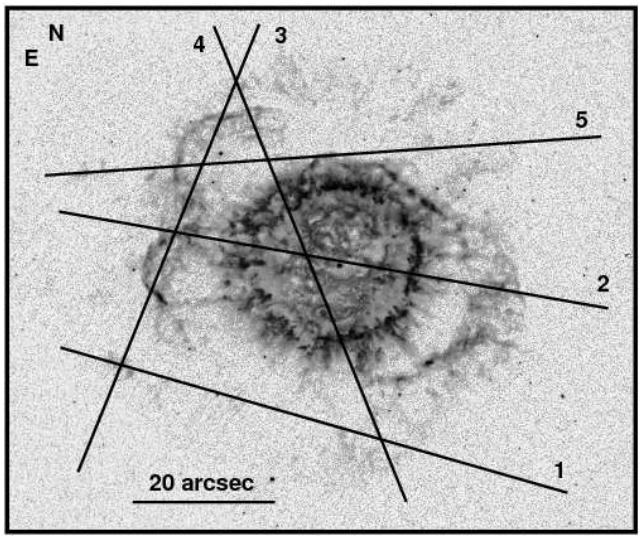}
\caption{
Slits positions used in the long-slit echelle spectroscopy of NGC\,6369 
superimposed upon the \emph{HST} WFPC2-WF3 [N~{\sc ii}] image.  
}
\label{slits.img}
\end{center}
\end{figure}

The ratio maps in Fig.~\ref{HST.img} reveal that the bright rim 
has high excitation, with the lowest [N~{\sc ii}]/H$\alpha$ and 
[N~{\sc ii}]/[O~{\sc iii}] values at radial distances 8\arcsec--\,12\arcsec, 
but the regions within this shell, closer to the central star, have 
anomalously low ionization.  
At the location of the ring of knots, at $\simeq$12\farcs4 from the central 
star, the ``cometary'' features are highlighted in the 
[N~{\sc ii}]/H$\alpha$ and [O~{\sc iii}]/H$\alpha$ ratio maps 
(Fig.~\ref{HST.img}), showing high ratio values in the former and low 
ratio values in the latter.  
The tails of these knots define an 
annular region with outer radius $\sim$18\arcsec.

The peculiar extensions of the inner shell of NGC\,6369 
\citep{SCM92} along PA$\simeq$63$\degr$ are best seen in the 
H$\alpha$ and [N~{\sc ii}] images.  
The western extension is reminiscent of a lobe or a large {\it ``ansa''}, 
whilst the eastern one displays a complex morphology best described as a 
bifurcated structure.  
The ratio maps indicate that these features have low excitation, with 
values of [N~{\sc ii}]/[O~{\sc iii}] up to 20 times larger than those 
in the higher excitation regions of the inner shell, and up to twice those 
of the ring of knots at the rim 
of the inner shell.  

\begin{figure*}
\begin{center}
\includegraphics[width=1.0\columnwidth]{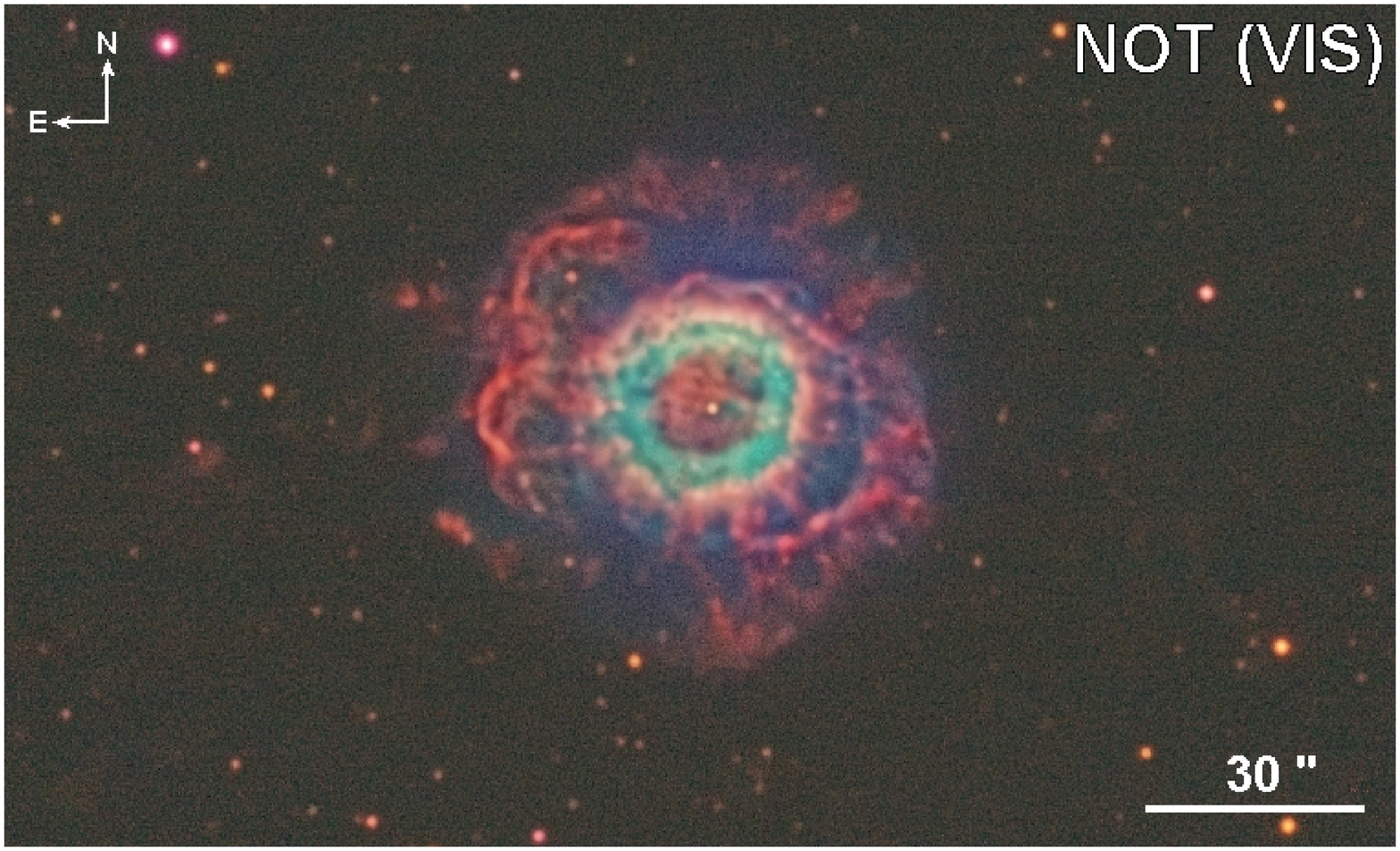}
\includegraphics[width=1.0\columnwidth]{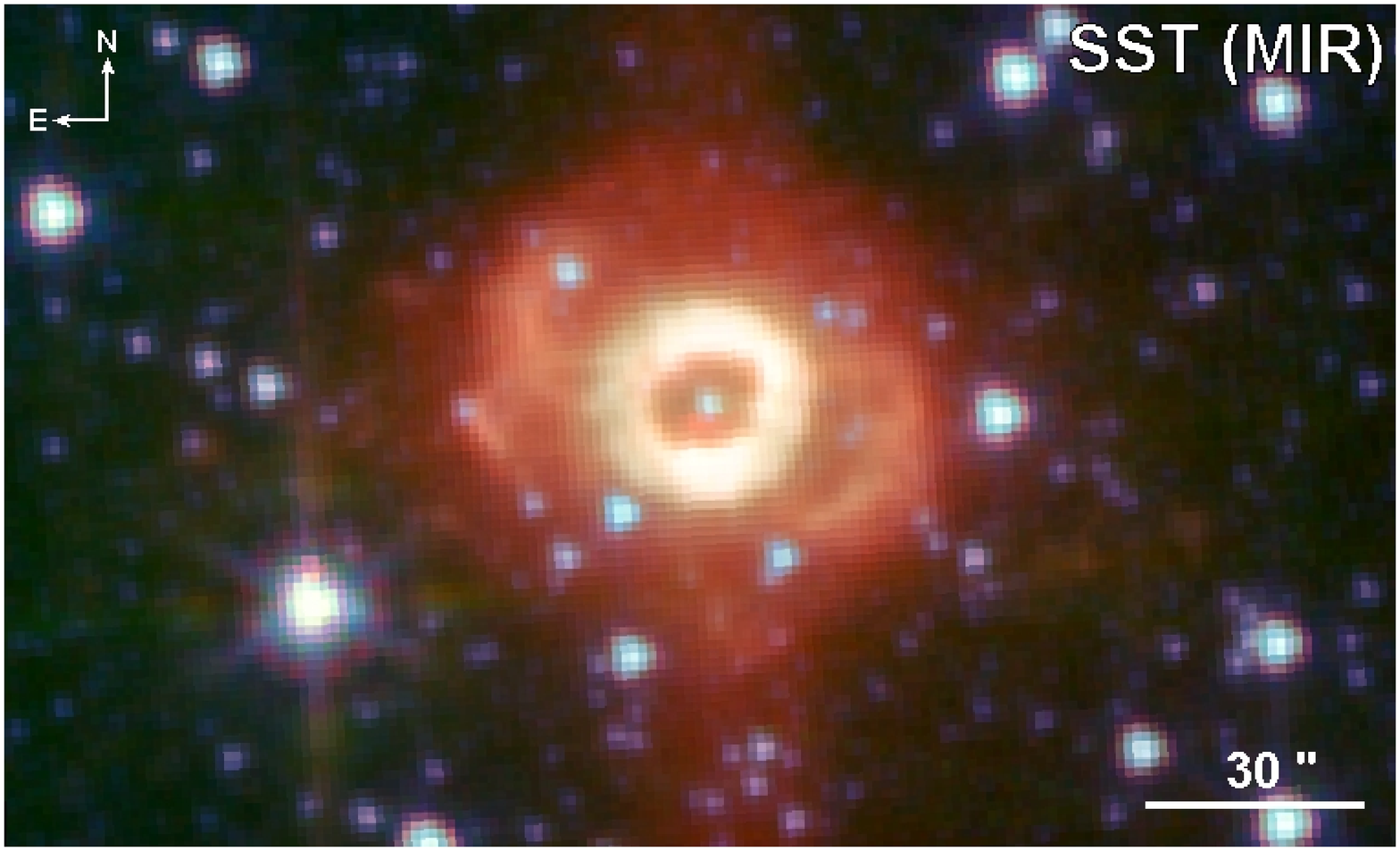}
\caption{
({\it left}) 
Colour-composite NOT [O~{\sc iii}] (blue), H$\alpha$ (green), and 
[N~{\sc ii}] (red), and 
({\it right}) \emph{Spitzer} IRAC 4.5 $\mu$m (blue), 5.8 $\mu$m 
(green), and 8.0 $\mu$m (red) pictures of NGC\,6369. 
We note that the stretch of the NOT optical images has been chosen to highlight the faint emission from the knots 
and condensations outside the ionised nebular shell.  
We also note that the north-south band of 8.0 $\mu$m emission 
is an instrumental artifact.  
}
\label{ext_emission}
\end{center}
\end{figure*}

The envelope is barely detected in [O~{\sc iii}], and only a little brighter in H$\alpha$.  
The [N~{\sc ii}] emission is brighter and reveals it to be composed of an ensemble of knots, blobs, 
and filaments.  
The appearance of these knots and their tails is less 
defined than those in the inner shell.   
Overall, the emission is mostly distributed within a fragmentary annular 
region with inner and outer radii 25\arcsec--\,34\arcsec, as illustrated 
by the NOT composite-colour picture of NGC\,6369 presented in 
Figure~\ref{ext_emission}-{\it left}.

\subsubsection{Near-infrared morphology}

The morphology of the bright inner shell and envelope of NGC\,6369 in 
the near-IR Br$\gamma$ line is similar to that in H$\alpha$ (Fig.~\ref{H2.img}-{\it right}).  
The lack of differential extinction between the optical and near-IR lines of H~{\sc i} suggests little dust content in the nebula along the line of sight. 
By contrast, the image in the near-IR H$_2$ line 
(Fig.~\ref{H2.img}-{\it left}) offers a very different 
view of the main shell of NGC\,6369.  
The region interior to the bright inner shell, which showed anomalously 
low excitation emission, is revealed to be filled with H$_2$ emission.  
This region is then encompassed by two concentric annular regions, 
the first one where the H$_2$ emission is very low, and 
the second one, with radius $\sim$13\arcsec, where we find bright 
H$_2$ emission.  
The interior annulus is coincident with regions of high excitation 
in the [N~{\sc ii}]/H$\alpha$ and [N~{\sc ii}]/[O~{\sc iii}] ratio 
maps, whilst the exterior, H$_2$-bright annulus is coincident with 
the rim of the inner shell defined by the heads of the [N~{\sc ii}] 
``cometary'' knots.  
This provides evidence that the [N~{\sc ii}] knots and filaments are 
associated with H$_2$ emission, i.e., they include significant amounts 
of neutral material.  

The extensions of the inner shell are not detected in H$_2$, but the 
northwest quadrant of the envelope and some bright knots exhibit 
H$_2$ emission.  
Surprisingly, the radius of the northwest arc detected in H$_2$ is 
$\sim$25\arcsec, i.e., this H$_2$ emission is interior to the annulus 
of [N~{\sc ii}] emission in the envelope.  
As for the H$_2$ knots, some (e.g., K3 and K5) have easily identifiable
[N~{\sc ii}] counterparts, while for some others (e.g., K2 and K4) the 
identification is hampered by their spatial coincidence with the 
[N~{\sc ii}]-bright eastern and western extensions of the main shell.  
Among the H$_2$ knots detected in the envelope, the morphology of knots K2 
and K3 is quite noticeable, with a bright head and two tails pointing 
outwards from the central star.

\subsubsection{Mid-infrared morphology}

The mid-IR morphology of NGC\,6369 is illustrated by the ``unsharp'' 
\emph{Spitzer} IRAC colour-composite picture presented in 
Figure~\ref{ext_emission}-{\it right}. 
In this picture, the bright optical shell is the brightest feature, 
whereas the outer envelope and east and west extensions are fainter 
but still noticeable.  
Despite the reduced spatial resolution, the mid-IR morphology 
of NGC\,6369 is consistent with that observed in the \emph{HST} 
(Fig.~\ref{HST.img}) and NOT (Fig~\ref{ext_emission}-{\it left}) 
optical images.  
The reddening of the eastern and western extension of 
the main shell and outer envelope in this colour-composite IRAC picture 
implies a higher relative contribution of 8.0 $\mu$m emission at these 
morphological components.  
This is further evidenced by the contours in Figure~\ref{irac_maps};  
the inner ring is intense in all of the IRAC images, whereas 
the surface brightness of the envelope becomes relatively brighter 
at longer wavelengths, in the 5.8 and 8.0 $\mu$m bands.  

\begin{figure*}
\begin{center}
\includegraphics[width=\columnwidth]{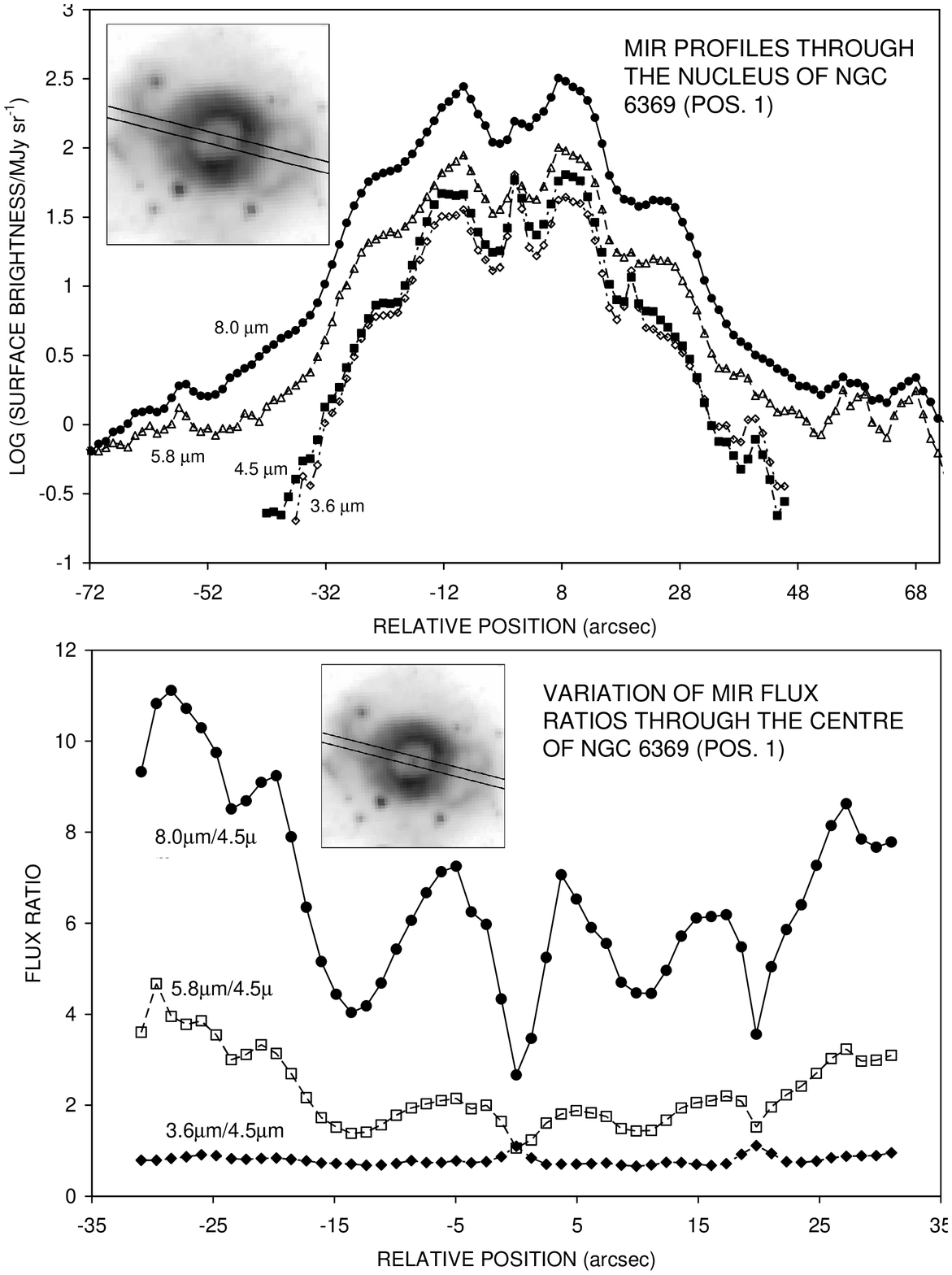}
\includegraphics[width=\columnwidth]{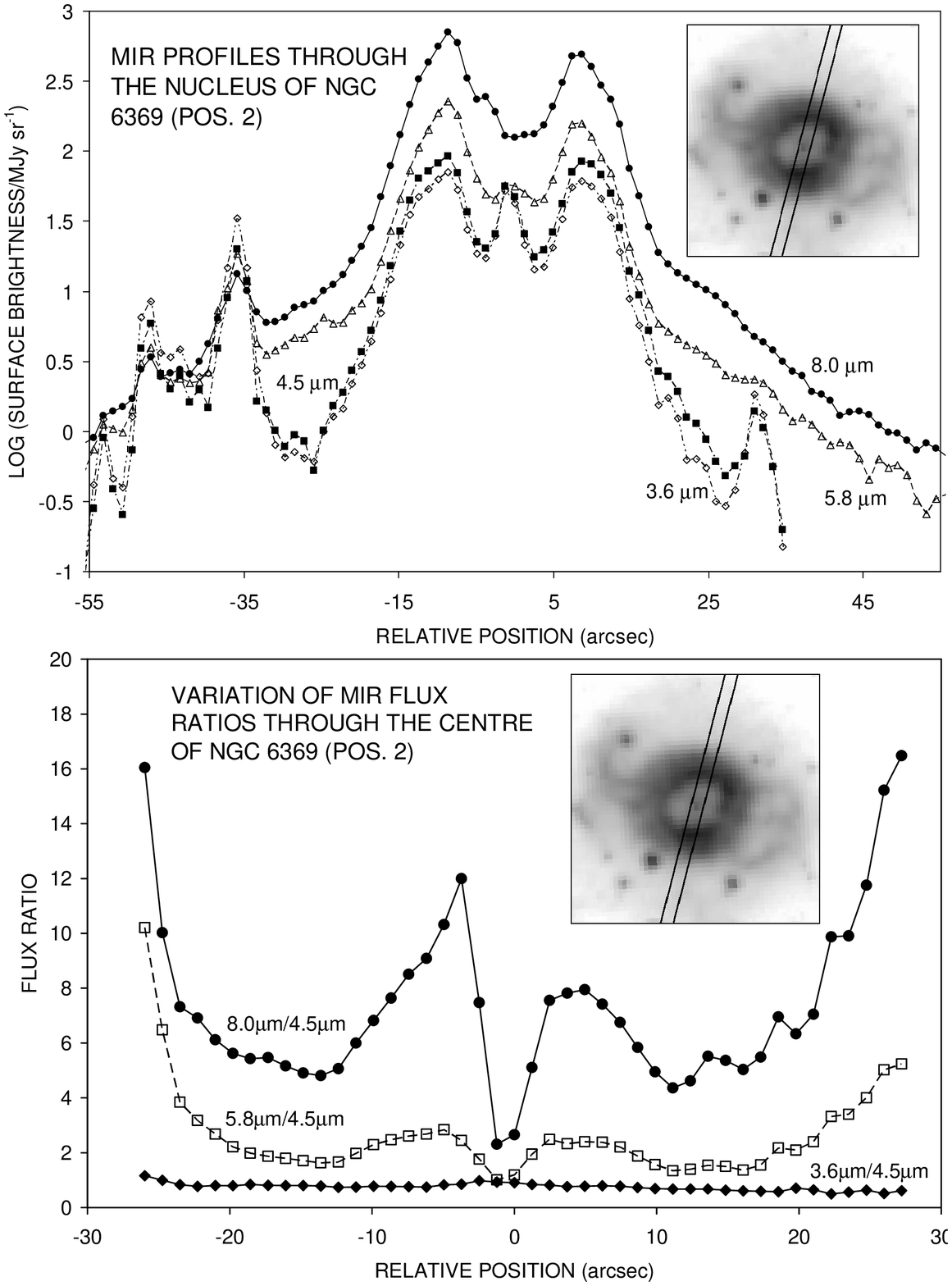}
\caption{
MIR IRAC profiles of NGC\,6369 along approximate E-W ({\it left}) and N-S 
({\it right}) axes through its central star as indicated in the inserted 
images. 
The lower panel shows flux ratios for the central region of the source, 
where 4.5 $\mu$m emission has a reasonably high S/N. 
}
\label{prof.plot}
\end{center}
\end{figure*}

The mid-IR surface-brightness profiles extracted from 
background-corrected IRAC images through the minor 
and major axes of the nebula (Figure~\ref{prof.plot}) 
confirm this result.  
The shape of the surface brightness profile interior to the bright 
main shell is similar in all of the channels, with flux increasing 
with wavelength.  
By contrast, the 3.6 and 4.5 $\mu$m surface brightness profiles 
fall-off sharply at $\sim$25\arcsec, whereas the emission at 
5.8 and 8.0 $\mu$m show broad skirts up to $\sim$70\arcsec.  
This becomes in a sharp increase of the 5.8$\mu$m/4.5$\mu$m and 
8.0$\mu$m/4.5$\mu$m ratios at radial distances greater than 
20\arcsec\ (Fig.~\ref{prof.plot}).  
We also note that the 5.8$\mu$m/4.5$\mu$m and 8.0$\mu$m/4.5$\mu$m 
ratios take minimum values at a radial distance $\sim$15\arcsec, 
i.e., the position of the bright [N~{\sc ii}] and H$_2$ ring.

\subsection{The extended emission}\label{ext_sec}

The \emph{HST} WF3 images (Fig.~\ref{HST.img}) hint the presence of 
irregular condensations outside the envelope of NGC\,6369, towards 
the east and close to the edge of the WF3 FoV.  
The NOT [N~{\sc ii}] image (Figure~\ref{ext_emission}-{\it left}), which 
covers a greater FoV than the \emph{HST} images, and very especially the 
WHT H$_2$ image (Figure.~\ref{H2.img}-{\it left}) reveal patches of diffuse 
emission and blobs, as marked by arrows in the H$_2$ image, well beyond the main ionised nebula, up to $\sim$70\arcsec\ from the 
central star in [N~{\sc ii}] and $\sim$80\arcsec\ in H$_2$.  
These [N~{\sc ii}] and H$_2$ blobs are distributed 
irregularly, mostly along the east-west direction, 
with some hints of point-symmetry.  
These blobs are also detected in the \emph{Spitzer} IRAC 8 $\mu$m image 
(Figure~\ref{ext_emission}-{\it right}), especially those that are well 
beyond the 8 $\mu$m bright envelope.

\section{Mid-IR spectroscopy}\label{mid_ir_sec}

The \emph{ISO} spectrum of NGC\,6369 in the 
2.4--4.1 $\mu$m spectral range shown in Figure~\ref{ISO} reveals the 
prevalence of emission lines, with Br$\alpha$ being the most intense 
and relatively weak or negligible emission of the 3.3 $\mu$m PAH band.  
There is also evidence for the H$_2$ 1$-$0 Q(3) $\lambda$2.424 $\mu$m 
and H$_2$ 1$-$0 Q(7) $\lambda$2.500 $\mu$m lines, which is 
consistent with the detection of H$_2$ emission in our WHT LIRIS 
images and at 17.036 $\mu$m \citep{PB-S08}.  
Although the spectrum remains relatively noisy, and some possible 
transitions are unidentified, we see evidence for Pf$\delta$ and 
Pf$\gamma$, as well as a couple of Humphrey transitions.  

The longer wavelength \emph{ISO} spectra are relatively noisy, precluding 
reliable analysis in the range $\Delta\lambda\sim$4.1--10 $\mu$m, although 
the enhancements in emission at 6.2 and 7.8 $\mu$m suggests that PAH band 
features are present.  
The \emph{Spitzer} IRS spectra (Figure~\ref{IRS}) show more 
convincing evidence for these PAH components.  
The shorter wavelength SL2 spectrum (Fig.~\ref{IRS}-{\it left}) 
shows evidence for the H$_2$ 0$-$0 S(7) $\lambda$5.511 $\mu$m 
line, but the spectrum is, otherwise, dominated by four PAH 
bands at 6.2, 6.9, 7.8, and 8.6 $\mu$m, and an underlying 
continuum which raises towards longer wavelengths and that 
can be attributable to cool dust emission.

The same situation applies at longer wavelengths as well 
(Fig.~\ref{IRS}-{\it right}), where the flux is dominated by 
PAH band emission yet again, with features at 7.8, 8.6, 11.3, 
12.7 and 14.2 $\mu$m contributing a significant fraction of 
the total emission.
We also see evidence for strong transitions due to the [Ar~{\sc iii}] 
$\lambda$8.991 $\mu$m and [S~{\sc iv}] $\lambda$10.51 $\mu$m lines, and 
the possible contribution of the Hu$\alpha$ $\lambda$12.372 $\mu$m 
line.  

One final point is related to the relative intensities of the PAH bands 
themselves. 
The 7.8 $\mu$m feature arises primarily through C-C stretching modes, whilst 
the 11.3 $\mu$m band is a consequence of C-H out-of-plane bending modes.  
This would normally lead to I(7.8$\mu$m)/I(11.3$\mu$m)$\approx$1.3 for 
neutral PAHs, but radiation fields in PNe are capable of appreciably 
ionising the PAH molecules, reducing the strength of the 11.3 $\mu$m 
feature and leading to I(7.8$\mu$m)/I(11.3$\mu$m)$\approx$12 for fully 
ionised PAHs \citep{LD01}.  
It is thus puzzling that the I(7.8$\mu$m)/I(11.3$\mu$m) ratio appears 
significantly less than unity (Fig.~\ref{IRS}-{\it right}) at any 
location of the nebula.  

\begin{figure*}
\centering
\includegraphics[height=5in]{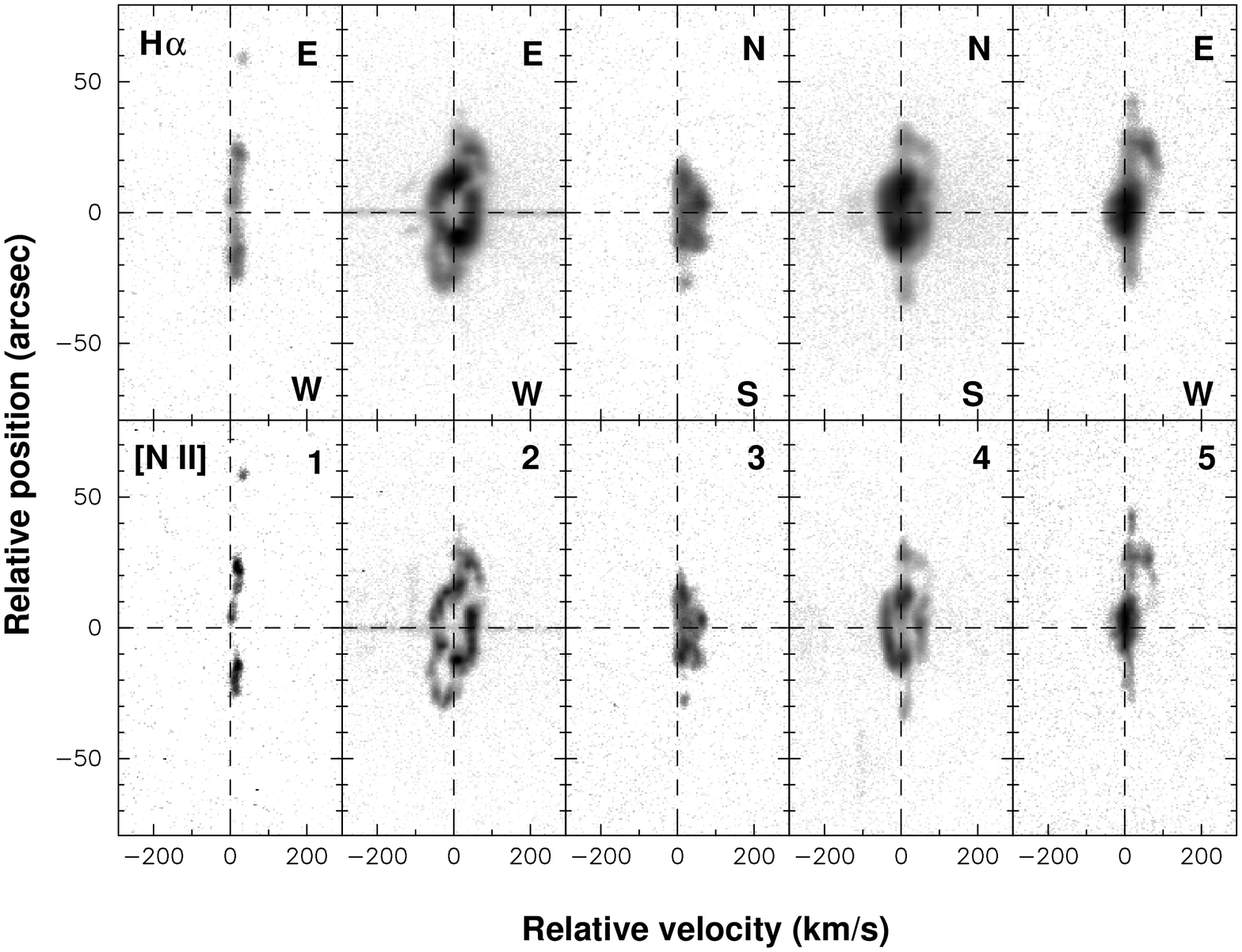}
  \caption{
({\it top}) H$\alpha$ and ({\it bottom}) [N~{\sc ii}] $\lambda$6583 
position-velocity (PV) maps of NGC\,6369 for slits positions \#1 to 
\#5.  
The slit orientation is labeled on the upper panels.  
The relative position is measured with respect to the horizontal dashed lines 
that mark the midpoint of the most intense central emission for slits \#1, \#3, 
\#4, and \#5, and the central star for slit \#2.  
The vertical dashed lines correspond to the systemic radial 
velocity ($V_{\rm LSR}=-89.5\pm1$\,km\,s$^{-1}$).
}
\label{PV.img}
\end{figure*}

\subsection{Interpreting the mid-IR images of NGC\,6369}

The mid-IR spectroscopy of NGC\,6369 described above allows us to 
discuss the nature of the emission detected in the IRAC bands.  
The \emph{ISO} data show convincingly that the emission in the IRAC 3.6 
and 4.5 $\mu$m bands is largely dominated by H~{\sc i} lines, with the 
Br$\alpha$ $\lambda$4.052 $\mu$m line in the 4.5 $\mu$m IRAC band being 
the brightest line.  
The contribution to these bands from Bremsstrahlung and 
the 3.3 $\mu$m PAH band is very small or negligible.  
The contribution from other ionic lines in the 4.1-5.1 $\mu$m 
spectral range \citep[e.g., {[Mg~{\sc iv}] $\lambda$4.486 $\mu$m, 
[Ar~{\sc vi}] $\lambda$4.529 $\mu$m, and [K~{\sc iii}] 
$\lambda$4.618 $\mu$m,}][]{Bernard_etal01} is also possible, but 
the \emph{ISO} spectrum has too low S/N and the \emph{Spitzer} 
IRS SL2 spectrum missed this spectral range.  
Meanwhile, the emission in the 5.8 and 8.0 $\mu$m IRAC channels is dominated by 
the 6.2, 6.9, 7.8, and 8.6 $\mu$m PAH bands, the [Ar~{\sc iii}] $\lambda$8.991 
$\mu$m emission line, and an underlying continuum from cool dust which raises 
towards longer wavelengths.  

The close resemblance between the images and surface brightness 
profiles in the 3.6 and 4.5 $\mu$m IRAC channels is due to the 
prominence of ionic and H~{\sc i} transitions in these two bands.  
The higher relative emission in the 5.8 and 8 $\mu$m bands at large radial 
distances implies that molecular material and cool dust, responsible of 
PAH bands and continuum emission, are important at the eastern and western 
extensions of the inner shell and at the envelope, whilst ionic lines 
dominates the emission from the bright inner shell.  
The relative increase of 5.8 and 8.0 $\mu$m, [N~{\sc ii}], and H$_2$ 
emission at the interior of the bright inner shell most likely reveal 
an opening of the inner shell at its polar caps, so that the emission 
detected at this region comes from regions far from the nebular center, 
above and below the bright inner shell.  

Finally, we noted in \S\ref{ext_sec} that the blobs and knots of diffuse 
emission outside the main nebular shells are detected in the optical 
[N~{\sc ii}], near-IR H$_2$, and IRAC 8 $\mu$m images.  
Since these blobs present both emission from ionised (N$^+$) and 
molecular (H$_2$) material, and the IRAC 8 $\mu$m band of the main 
nebula is dominated by PAH bands and dust continuum, the origin of 
the emission in the IRAC 8 $\mu$m band may be multi-fold.  
The [Ar~{\sc iii}] $\lambda$8.991 $\mu$m line is the brightest ionic line in 
this band, but the ionization potential of Ar$^{+}$ (IP=27.6 eV) is higher 
than this of N (IP=14.5 eV) and thus Ar$^{++}$ may be not present in these 
regions.  
Indeed, the lack of [O~{\sc iii}] emission, whose IP(O$^{+}$) 
is 35.1 eV, indicates that only low-excitation species are 
present in these regions.  
Since no emission from these blobs is detected in the IRAC 5.8 $\mu$m 
image, it seems unlikely that PAH bands or dust continuum produce the 
emission detected in the IRAC 8 $\mu$m image, although we concede that 
the levels of dust continuum and PAH emission in the IRAC 5.8 $\mu$m 
may be lower than in the IRAC 8 $\mu$m band.  
Given the previous difficulties for ionic, PAH or continuum emission and that 
near-IR H$_2$ 1$-$0 S(1) emission is detected, we favor the H$_2$ 0$-$0 S(4) 
$\lambda$8.0251 $\mu$m emission line as the responsible for the emission from 
these blobs in the IRAC 8 $\mu$m image.

\section{Discussion}\label{sec_dis}

\subsection{Spatio-kinematical modelling of the inner shell of NGC\,6369}
\label{kin_model}

Figure~\ref{PV.img} displays the individual H$\alpha$ and [N~{\sc ii}] 
position-velocity (PV) maps of NGC\,6369 obtained from the five slits 
positions illustrated in Fig.~\ref{slits.img}.  
Overall, the H$\alpha$ and [N~{\sc ii}] lines show similar 
kinematical features, but these are best seen in [N~{\sc ii}] 
given that its thermal broadening is smaller than 
that of the H$\alpha$ line.  
The radial velocity of the source with respect to the Local Standard of 
Rest was determined to have a value $V_{\rm LSR}=-89.5\pm1.0$\,km\,s$^{-1}$, 
based on the mean expansion velocity at the position of the central star.  
This result is in excellent agreement with the value of 
$V_{\rm LSR}\simeq-90\pm6.5$\,km\,s$^{-1}$ reported by 
\citet{Schneider_etal83}.  
 
The line shapes of the echellogram at slit position \#2, which cuts 
the nebula along the east and west extensions, is closely similar to that 
presented by \citet{SL06}: 
the east and west extensions are detected as red- and blue-shifted 
loops, respectively, and their shape is typical of the emission in 
bipolar outflows.  
The structure of the H$\alpha$ line may give the false impression that 
the inner shell is a closed ellipsoid, but the sharper view offered by 
the [N~{\sc ii}] line reveals that the loops of the east and west 
extensions emanate from gaps at the tips of the inner shell.  

The eastern extension is mapped in more detail by the slits at 
positions \#3, \#4, and \#5.  
In the echellograms at positions \#4 and \#5, this extension is detected 
as a red-shifted loop, as expected for a blister-like structure.  
However, the echellogram at position \#2 reveals a waist and two 
open loops that are indicative of a more complex structure.    

Previous three-dimensional modeling of NGC\,6369 \citep{Monteiro_etal04,SL06} 
concur that this nebula has a bipolar structure, although the details make 
them differ significantly.  
\citet{Monteiro_etal04} obtained spatially-resolved information of 
the physical conditions within the nebula that, in conjunction with 
a photo-ionization model, allowed them to synthesize narrow-band 
images.  
In order to match the morphology observed in several emission 
lines, they adopted for NGC\,6369 a clumpy hourglass structure with a thick 
equatorial belt and two symmetrically located caps of emission whose 
orientation is tilted by 30$\degr$ from the symmetry axis of the main nebula.  
On the other hand, \citet{SL06} used a single high-dispersion 
long-slit spectrum of NGC\,6369 obtained along its major axis 
to illustrate the capabilities of {\it ``SHAPE''}, an 
interactive 3-D modeling tool.  
Using a Hubble-type ($v{\propto}r$) expansion law for NGC\,6369, as it is 
typically assumed in PNe \citep[e.g.,][]{SU85,GMS96,Sab_etal00,Corradi04}, 
they concluded that there is no equatorial waist in NGC\,6369, but its 
three-dimensional structure can rather be described by a partial spheroidal 
shell with opened tips at its major axis from where two fainter blisters protrude as the east end west extensions.

\begin{figure}
\begin{center}
\includegraphics[width=\columnwidth]{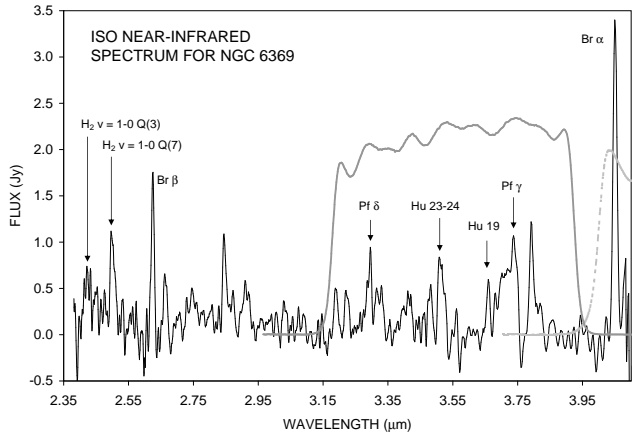}
\caption{
\emph{ISO} spectrum of NGC\,6369 extending between 2.4 and 4.1 $\mu$m, 
together with the ranges corresponding to the 3.6 $\mu$m (solid 
profile) and 4.5 $\mu$m (dashed profile) \emph{Spitzer} IRAC 
channels. 
}
\label{ISO}
\end{center}
\end{figure}

Our more detailed spatial coverage of the kinematics of NGC\,6369 allows us 
to refine \citet{SL06}'s model and to investigate the spatio-kinematical 
structure of the nebular envelope.  
Using the latest version of the code \emph{SHAPE} \citep[version 
10;][]{ST11}, we have constructed a spatio-kinematical model for 
NGC\,6369 which fits its narrow-band images and PV maps.  
The quality of the fit can be judged in Figure~\ref{model1} that presents 
a comparison between the synthetic morphology and kinematics from the model 
and those observed in NGC\,6369.  
The synthetic image of NGC\,6369 (right image of the upper panel in 
Fig.~\ref{model1}) is well matched with its \emph{HST} H$\alpha$ 
image (left image).  
Similarly, the kinematics of NGC\,6369 are fairly reproduced as shown in 
the lower panel of Fig.~\ref{model1} which presents the PV for slits \#2 
to \#5 accompanied by the corresponding PV map extracted from the model. 
Our best fit model, illustrated in Figure~\ref{model2}, is composed of: 
\\
$\bullet$ an equatorial barrel with expansion velocity $\simeq$55 km~s$^{-1}$ and 
whose axis of symmetry is inclined by $\sim$50\degr\ with respect to the 
line of sight (RLoS), 
\\
$\bullet$ a southern lobe approaching to us with an inclination angle 
$\sim$40\degr\ (RLoS) and polar expansion $\simeq$60 km~s$^{-1}$, 
and 
\\
$\bullet$ two northern lobes with inclination angles $\sim$115\degr\ and 
$\sim$127\degr\ receding from us with polar expansion velocities 
$\simeq$77 and $\simeq$84 km~s$^{-1}$, respectively. \\
In this way, the physical structure of NGC\,6369 is reminiscent 
of other barrel-shaped PNe with polar extensions that result in a 
spindle shape \citep[e.g., NGC\,3918,][]{Corradi_etal99}.

\begin{figure*}
\begin{center}
\includegraphics[height=6cm]{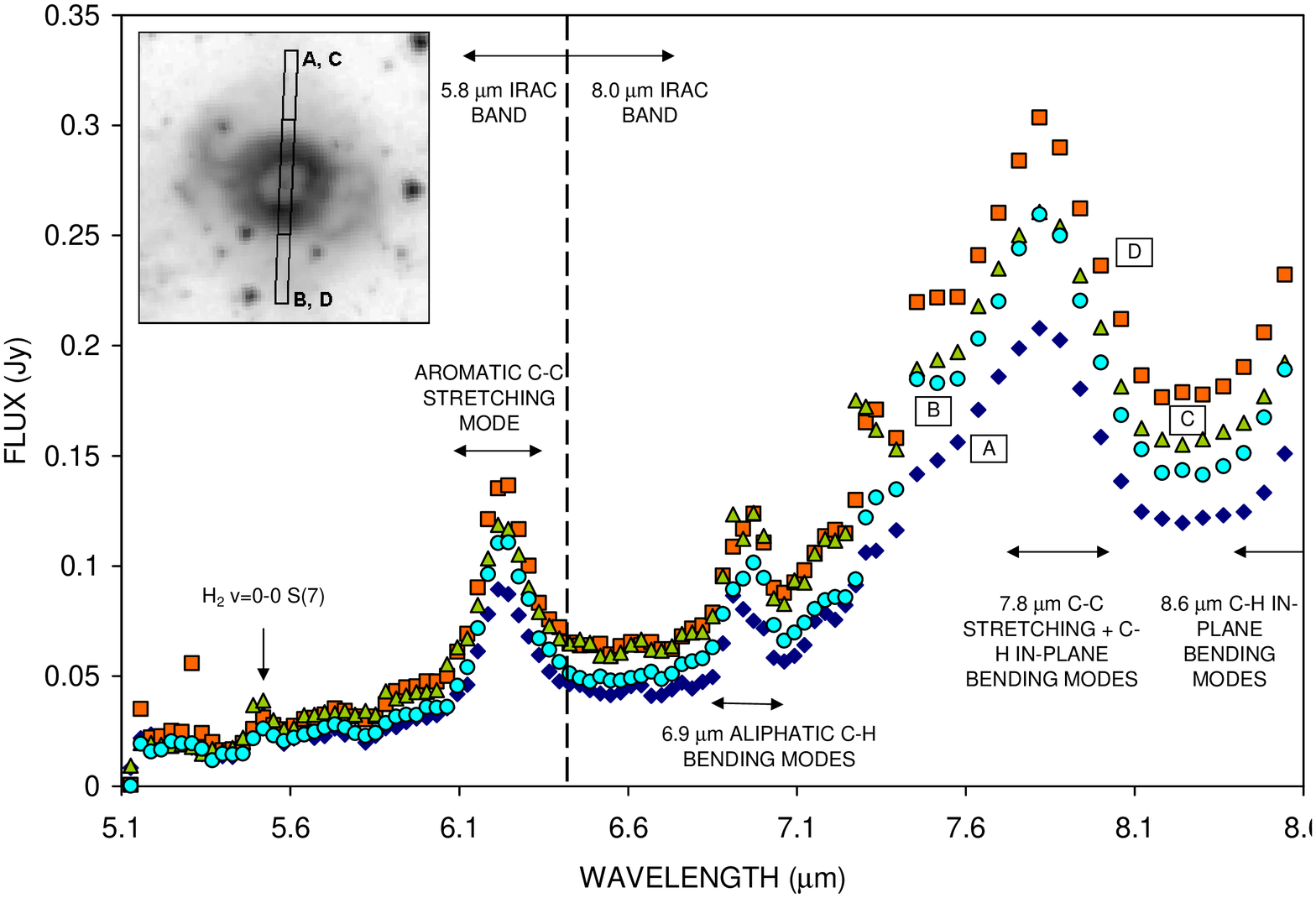}
\includegraphics[height=6cm]{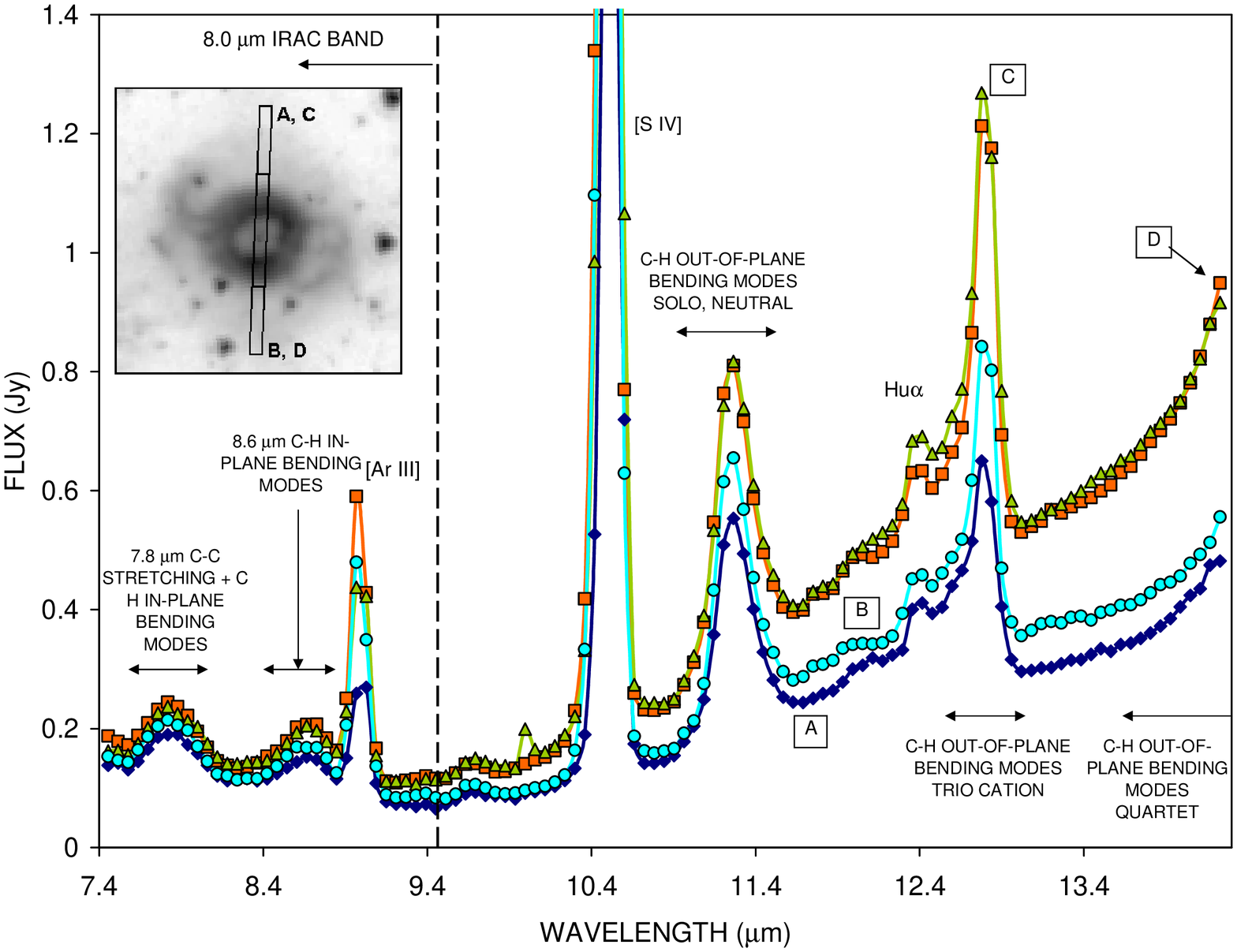}
\caption{
\emph{Spitzer} IRS spectrum of NGC\,6369 in the SL2 short ({\it left}) and SL1 
long ({\it right}) wavelength ranges taken at the four overlapping positions 
shown in the inset: A (dark blue diamonds), B (cyan open dots), C (green 
triangles), and D (red squares). 
Positions A and C cover the inner shell and north portion of the 
envelope, whereas positions B and D also cover the inner shell, but 
the south portion of the envelope.
Despite the (symmetric) displacement between positions B+D and A+C, the 
spectra are mostly consistent. 
The spectral range $\sim$ 7.3--7.6 $\mu$m corresponds at two overlapping 
spectral sequences, and results in this regime should be considered as 
unreliable.} 
\label{IRS}
\end{center}
\end{figure*}

According to this model and for the distance of 1,550 pc proposed by 
\citet{Monteiro_etal04}, the kinematical ages of the barrel and bipolar 
lobes are $\simeq$2,000~yrs and $\simeq$3,200~yrs, respectively.  
The kinematical age of the inner shell is thus smaller than that 
of the bipolar lobes, most likely indicating important dynamical 
effects that have made the kinematical and real ages to differ.  
These dynamical effects may also be revealed by the distorted eastern 
structure whose complex shape would arise from the interaction of an 
expanding, regular lobe with extended and irregular material along its 
direction. 
Alternatively, this bifurcated structure may arise from intrinsic multipolar collimated outflows as has 
been claimed in other PNe \citep{MSG96,Sahai00}.

\subsection{Kinematics of the envelope}

The emission from the outer envelope of NGC\,6369 is registered by the 
slits at positions \#1, \#3, \#4, and \#5, although the echellograms at 
positions \#3 and \#4 only cover a small fraction of this shell.  
In the echellograms at positions \#1 and \#5, the H$\alpha$ and [N~{\sc ii}] 
lines are extremely narrow, with $\rm FWHM\simeq32\,km\,s^{-1}$ and $\simeq20$ 
km~s$^{-1}$, respectively.  
We note, however, that the velocity centroid along these lines is not 
constant: the echellogram at position \#1 shows a weavy pattern (best 
seen in the [N~{\sc ii}] line), whereas at position \#5 displays an 
arc-like shape.  
We measure velocity differences up to 17 km~s$^{-1}$.  
We also note that emission is patchy, especially in the [N~{\sc ii}] 
line.  

Clearly the envelope is not inert, but the narrow and weavy/arced line shape 
does not comply with the expectations for an expanding shell, even if it had 
the low expansion velocity \citep{Meaburn_etal96,GVM98} presumed for the slow 
AGB wind.  
The knotty and fragmented appearance of the envelope of NGC\,6369 
indicate that the kinematics revealed by the echellograms is 
governed by the emission from individual knots, rather than by the 
emission from a continuous shell.  
This situation is reminiscent of many other PNe where bright low-ionization 
knots are embedded within a faint envelope around a bright inner shell, e.g., 
NGC\,2392 and NGC\,7354 \citep{O'Dell_etal2002,Contreras_etal2010}.  
As for these two PNe, the relatively low velocity differences measured 
for the knots of the envelope of NGC\,6369 would suggest that they lay 
at different latitudes with respect to the symmetry axis of the barrel, 
but mostly on a weavy disk-like or flattened structure, as for NGC\,2392 
and NGC\,7354.

\subsection{Extension of the photo-dissociation region}

The relative sizes of the regions dominated by ionized material and 
molecules and dust are illustrated in Figure~\ref{prof.han2h2i4i8} 
that displays the normalized surface brightness profiles in the 
\emph{HST} H$\alpha$ and [N~{\sc ii}], WHT LIRIS H$_2$, and 
\emph{Spitzer} IRAC 4.5 and 8.0 $\mu$m images.  
The comparison between these profiles confirms that the surface brightness 
profiles of the inner shell of NGC\,6369 of the optical H$\alpha$ emission 
line and mid-IR IRAC 4.5 and 8.0 $\mu$m bands are very similar in shape and 
extent.  
The mid-IR and optical profiles start to depart from each other 
at the edge of the inner shell: the emission from H$\alpha$ and 
[N~{\sc ii}] drops considerably in an annular region between 
20\arcsec\ and 25\arcsec\ (i.e., the inner regions of the 
envelope), while the emission in the IRAC 4.5 and 8.0 $\mu$m 
bands stays higher and the H$_2$ profile shows increased 
emission.  
On the contrary, on the 25\arcsec--35\arcsec\ outer rim of the envelope, 
the emission in the H$\alpha$ and [N~{\sc ii}] lines is enhanced, while 
the emission in the IRAC 4.5 and 8.0 $\mu$m bands flattens and then 
decays smoothly.  
There is no H$_2$ emission in this range of radial distances.

\begin{figure*}
\begin{center}
\includegraphics[width=1.4\columnwidth]{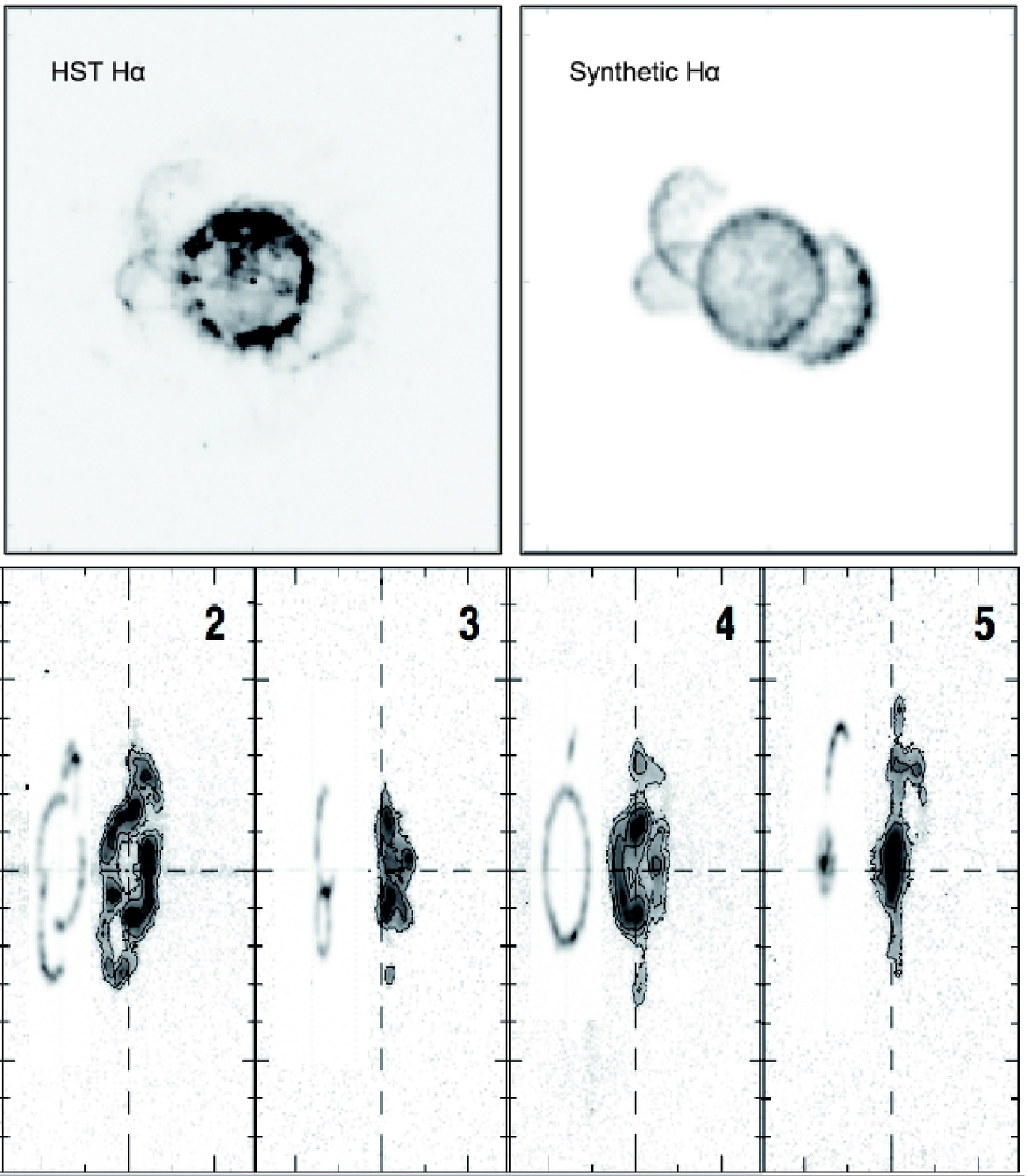}
\caption{
({\it top}) \emph{HST} H$\alpha$ ({\it left}) and synthetic H$\alpha$ 
({\it right}) images of NGC\,6369.  
({\it bottom}) PV maps from the model compared to the observed PV maps.  
The synthetic PV maps have been displaced towards higher velocities (to the 
left in the PV maps) to allow a fair comparison.  
}
\label{model1}
\end{center}
\end{figure*}

The comparison of the relative shapes and extents of these different 
surface brightness profiles is relevant to the understanding of the 
physical structure of NGC\,6369.  
Its inner shell is mostly ionized and thus the emission in the 
IRAC 4.5 and 8.0 $\mu$m bands is dominated by ionic emission.  
The shell is enclosed by a series of ``tadpole'' or ``cometary'' knots whose 
prominent low-excitation [N~{\sc ii}] and molecular H$_2$ emission at their 
heads result in bright  [N~{\sc ii}] and H$_2$ peaks at $\simeq$13\arcsec.  
The inner shell is surrounded by a region dominated by molecular 
material, the inner rim of the envelope between $\sim$17\arcsec\ 
and $\sim$25\arcsec.  
This region, outside the bright inner shell and beyond the annular 
region defined by the tails of the ``cometary'' knots, seems to be 
a real PDR.  
The relative brightening of the emission in the IRAC 4.5 and 8.0 $\mu$m 
bands at these radial distances is very likely due to molecular H$_2$ 
emission, although a contribution from continuum dust or PAH emissions 
is also very likely as noted in other PNe 
\citep[e.g.,][]{PR-L08,PR-L10,R-LP08}.  
Finally, the outer rim of the envelope is again dominated by ionic 
emission, but we note that low-excitation emission dominates.  
This may explain the flattening of the emission in the 
IRAC 4.5 and 8.0 $\mu$m emission due to an additional 
contribution of ionic emission.  

It is finally worth noting that sources with comparable He~{\sc ii} and 
H~{\sc i} Zanstra temperatures are likely to be ionisation bound and 
enveloped by a shell of neutral material. 
This leads to a correlation between the Zanstra ratio 
T$_Z$(He~{\sc ii})/T$_Z$(H~{\sc i}), and mid-IR extent 
outside of the ionised shell \citep[see e.g.,][]{Q-MPR-L11}.  
NGC\,6369 seems to be a peculiar case of this correlation: 
whereas the emission from molecular material causes a larger dimension of its 
inner shell at mid-IR wavelenghts, the presence of ionized material in its 
singular envelope results in very similar final optical and mid-IR dimensions.

\subsection{Shock excitation?}

In \S\ref{neb_morp_sec} we noted the large jump in the [N~{\sc ii}] 
to [O~{\sc iii}] ratio at the filamentary structures that enclose 
the eastern and western extensions of the inner shell of NGC\,6369.  
Given that these structures are at the tips of bubble-like structures 
that may arise from bipolar outflows, it is very likely that the unusually 
high [N~{\sc ii}]/[O~{\sc iii}] ratios stem from shock interactions of these structures with 
surrounding material.  
To investigate the structure of these features in more detail, 
we show in Figure~\ref{fig.shocks} a radial profile across a 
particularly bright filament of the eastern extension extracted 
from the \emph{HST} H$\alpha$, [N~{\sc ii}], and [O~{\sc iii}] 
images and ratio maps. 

The comparison of the surface brightness profiles indicates that the 
peak of all these three emission lines are almost coincident, but the 
spatial profile of the [O~{\sc iii}] line is noticeable broader.  
The inner shoulder of this profile, closer to the central star 
(negative offsets in Fig.~\ref{fig.shocks}), can be explained 
as the result of a higher photo-excitation.  
The shoulder of the line farther away from the nebula, however, cannot 
be explained in this way and it very likely reveals the presence of a 
forward shock propagating onto a lower density medium.  
The [O~{\sc iii}]/H$\alpha$ ratio profile in the lower-panel of 
Fig.~\ref{fig.shocks} allows us to derive a quantitative value 
for the spatial extent of this post-shock region, $\simeq1\farcs2$.  
At the distance of 1,550 pc, this corresponds to 
$\sim$3$\times$10$^{16}$ cm, comparable to the 
post-shock cooling widths determined from modelling 
of shocks \citep[e.g.,][]{Dopita77,SMcK79,Pittard_etal05}.

\subsection{The external condensations of NGC\,6369}

There is a growing number of PNe in the literature that show knots or 
blobs of emission well separated from their main nebular shells that 
do not form a halo.  
In some cases, the kinematics of these knots imply high-velocity collimated 
outflows, e.g., Fleming~1 \citep{PLML96} or MyCn\,18 \citep{Redman_etal00}, 
while in some others, e.g., IC\,4634 \citep{Guerrero_etal08}, the velocity 
of the knots is close to the systemic velocity, may be due to motions close 
to the plane of the sky.  

As for NGC\,6369, we note that one of the blobs located outside its 
inner shell and envelope is detected in the H$\alpha$ and [N~{\sc ii}] 
emission lines at the slit position \#1 (Fig.~\ref{slits.img}).
This blob, at an offset of $\sim$70\arcsec \,from NGC\,6369 central 
star, has a relatively small velocity shift, $\sim$+17 km~s$^{-1}$, 
with respect to the systemic velocity, and its velocity width is 
small, $\rm FWHM\simeq26\,km\,s^{-1}$ in H$\alpha$ and $\simeq21$ 
km~s$^{-1}$ in [N~{\sc ii}].

The kinematical properties of this outer condensation make NGC\,6369 
closer to the case of the lower velocity knots of IC\,4634 than to a 
case of high-velocity collimated outflows.  
Meanwhile, the velocity shift, although small, would preclude that these 
outer condensations are part of an incomplete halo \citep[as, e.g., the 
case of IC\,4593,][]{Corradi_etal97}.  
The origin of these external condensations is uncertain, but they may probe earlier, point-symmetric ejections, or arise from shocks associated with the east-west that excite 
material surrounding the nebula.

\subsection{Physical structure of NGC\,6369}

The spatio-kinematical model of the inner shell of NGC\,6369, the kinematics 
of its envelope, and the intrincate distributions of ionized, molecular, and 
mid-IR emission can be used to obtain a coherent view of its structure.  
The inner shell and extensions of NGC\,6369 can be described as a tilted equatorial barrel with bipolar extensions. 

\begin{figure}
\begin{center}
\includegraphics[width=0.975\columnwidth]{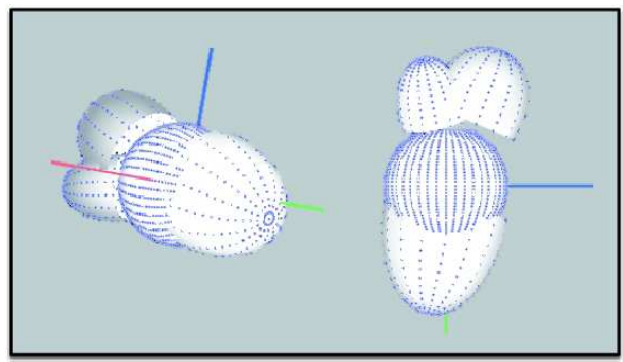}
\caption{
Main structures of NGC\,6369 represented in a 3D model. 
The left panel corresponds to the inclination and PA proposed. 
The right panel is the same structure but rotated to show a 
complementary view of the model. 
}
\label{model2}
\end{center}
\end{figure}

This inner shell is encircled by a ring of [N~{\sc ii}]-bright ``cometary'' 
knots.  
The envelope is formed by an ensemble of discrete low-ionization and 
molecular blobs and knots probably embedded within a much fainter, 
fully ionized shell.  
These knots seem to be distributed at slightly different latitudes 
with respect to the barrel symmetry axis, but mostly on a weavy 
disk-like or flattened structure.  
  
Near- and mid-IR H$_2$ emission encompasses the inner shell, revealing 
a real PDR that seems to be inside the envelope, a shell of ionized 
gas.  
It is unclear whether this PDR exists within the envelope, or whether this 
is a projection effect, so that the PDR is located at high nebular latitudes, 
above and below the envelope that would be a rather flat structure only 
present at low latitudes.

\begin{figure}
\begin{center}
\includegraphics[bb=18 240 570 570,width=0.99\columnwidth]{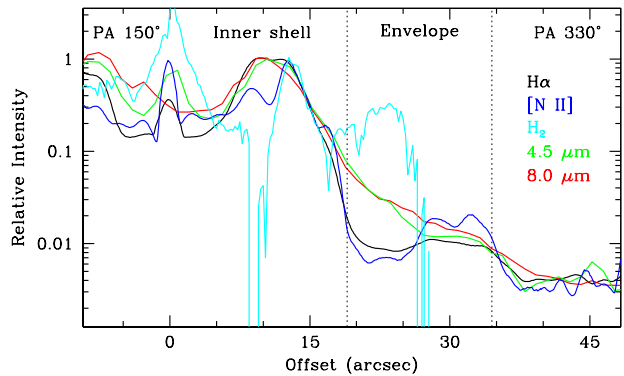}
\caption{
Surface brightness profiles of the H$\alpha$ (black), [N~{\sc ii}] 
(blue), WHT H$_2$ (cyan), IRAC 4.5 $\mu$m (green), and IRAC 8.0 $\mu$m (red) 
along PA 330\degr\ through the central star of NGC\,6369.  
The profiles have been normalized to the emission peak at the bright 
inner shell.  
The location of the inner shell end envelope edges are marked by 
vertical dotted lines.  
}
\label{prof.han2h2i4i8}
\end{center}
\end{figure}

\section{Conclusions}\label{sec_con}

Using narrow-band optical and near-IR images, broad-band mid-IR 
images, optical long-slit echelle spectra and mid-IR spectra, we 
have investigated the physical structure of the planetary nebula 
NGC\,6369.  
The inner shell and its east-west extensions can be modeled in terms of 
a tilted barrel of ionized gas and bipolar outflows at its tips, in 
broad agreement with previous observations of the source.  
The nebula is optically thick to H ionizing radiation, thus allowing 
the existence of a PDR that is confirmed by its H$_2$ emission and 
suggested by an additional contribution of dust continuum and PAH 
band emission to the IRAC images in the 5.8 and 8 $\mu$m bands.  
The envelope does not seem to be a real shell enclosing the inner 
shell, but rather a flattened structure at the equatorial regions 
of NGC\,6369.  
Molecular and low-excitation emission of [N~{\sc ii}] is also detected 
in small irregular condensations well outside the main nebular shells.  
Finally, we note that \emph{HST} images show the presence of 
``cometary'' knots at the rim of the inner shell whose appearance 
is similar to those of features in NGC\,2392 and NGC\,7354, among 
other PNe.

\begin{figure}
\begin{center}
\includegraphics[width=0.99\columnwidth]{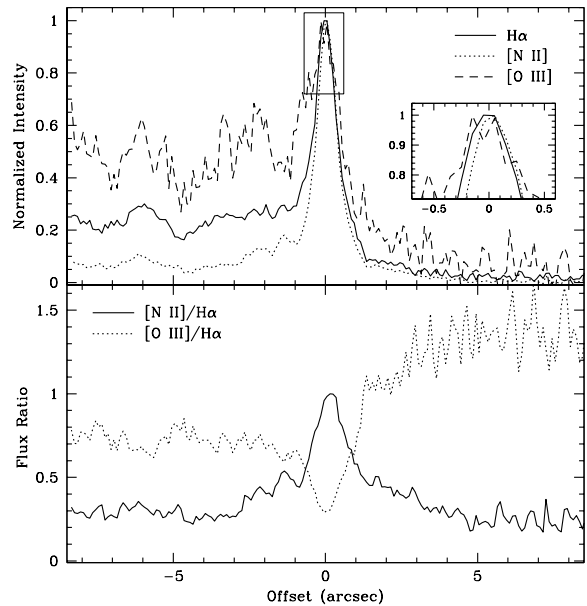}
\caption{
({\it top}) [N~{\sc ii}], H$\alpha$, and [O~{\sc iii}] surface 
brightness profiles through a likely shock structure in the eastern 
extension of the inner shell of NGC\,6369.  
Negative offsets are closer to the central star, while 
positive offsets corresponds to regions farther away 
from the main nebula.  
({\it bottom}) [N~{\sc ii}]/H$\alpha$ and [O~{\sc iii}]/H$\alpha$ 
ratio profiles.  
}
\label{fig.shocks}
\end{center}
\end{figure}

\section*{Acknowledgments}

GRL acknowledges support from CONACyT and PROMEP (Mexico).
MAG and GRL are partially funded by grant AYA2008-01934 of the 
Spanish Ministerio de Ciencia e Innovaci\'on (MICINN).
RV, MAG and GRL thank support by grant IN109509 (PAPIIT-DGAPA-UNAM).
We also would like to thank to the referee R. Corradi who made very excellent comments for the improvement of the paper.

Based on observations made with the Observatorio Astron\'omico Nacional at the Sierra de San Pedro M\'artir, OAN-SPM, wich is operated by the Instituto de Astronom\'{\i}a of the Universidad Nacional Aut\'onoma de M\' exico.

The William Herschel Telescope is operated on the island of La Palma by the Isaac Newton Group in the Spanish Observatorio del Roque de los Muchachos of the Instituto de Astrof\'{\i}sica de Canarias.

Based on observations made with the Nordic Optical Telescope, operated on the island of La Palma jointly by Denmark, Finland, Iceland, Norway, and Sweden, in the Spanish Observatorio del Roque de los Muchachos of the Instituto de Astrof\'{\i}sica de Canarias. 

This research has made use of the NASA/ IPAC Infrared Science Archive, which is operated by the Jet Propulsion Laboratory, California Institute of Technology, under contract with the National Aeronautics and Space Administration.

We would like to dedicate this paper in memory of our colleague and friend, Prof. John Peter Phillips, who recently passed away.

\bsp

\label{lastpage}

\end{document}